\documentclass[prx,aps,twocolumn,reprint,superscriptaddress,showpacs,floatfix,footinbib,longbibliography]{revtex4-1}
\usepackage{amsmath}
\usepackage{amssymb}
\usepackage{amsthm}
\usepackage{graphicx}
\usepackage{subfigure}
\usepackage{mathtools}
\usepackage{amsfonts}
\usepackage{float}

\newcommand{\be}{\begin{equation}}
\newcommand{\ee}{\end{equation}}
\newcommand{\bea}{\begin{eqnarray}}
\newcommand{\eea}{\end{eqnarray}}
\newcommand{\mbf}{\mathbf}
\newcommand{\mrm}{\mathrm}
\newcommand{\Rb}[1]{^{#1}\mathrm{Rb}}
\renewcommand{\vec}[1]{\mathbf{#1}}
\newcommand{\epstil}{\tilde{\epsilon}}
\newcommand{\kaptil}{\tilde{\kappa}}

\begin{document}
\title{Tunable-range, photon-mediated atomic interactions in multimode cavity QED}
\author{Varun D. Vaidya}
\affiliation{Department of Physics, Stanford University, Stanford, CA 94305}
\affiliation{Department of Applied Physics, Stanford University, Stanford, CA 94305}
\affiliation{E.~L.~Ginzton Laboratory, Stanford University, Stanford, CA 94305}
\author{Yudan Guo}
\affiliation{Department of Physics, Stanford University, Stanford, CA 94305}
\affiliation{E.~L.~Ginzton Laboratory, Stanford University, Stanford, CA 94305}
\author{Ronen M. Kroeze}
\affiliation{Department of Physics, Stanford University, Stanford, CA 94305}
\affiliation{E.~L.~Ginzton Laboratory, Stanford University, Stanford, CA 94305}
\author{Kyle E. Ballantine}
\affiliation{SUPA, School of Physics and Astronomy, University of St Andrews, St Andrews KY16 9SS UK}
\author{Alicia J. Koll\'{a}r}
\affiliation{Department of Applied Physics, Stanford University, Stanford, CA 94305}
\affiliation{E.~L.~Ginzton Laboratory, Stanford University, Stanford, CA 94305}
\author{\\Jonathan Keeling}
\affiliation{SUPA, School of Physics and Astronomy, University of St Andrews, St Andrews KY16 9SS UK}
\author{Benjamin L. Lev}
\affiliation{Department of Physics, Stanford University, Stanford, CA 94305}
\affiliation{Department of Applied Physics, Stanford University, Stanford, CA 94305}
\affiliation{E.~L.~Ginzton Laboratory, Stanford University, Stanford, CA 94305}
\date{\today}

\begin{abstract}
Optical cavity QED provides a platform with which to explore quantum many-body physics in driven-dissipative systems.  Single-mode cavities provide strong, infinite-range photon-mediated interactions among intracavity atoms.  However, these global all-to-all couplings are limiting from the perspective of exploring quantum many-body physics beyond the mean-field approximation.  The present work demonstrates that local couplings can be created using multimode cavity QED.  This is established through measurements of the threshold of a superradiant, self-organization phase transition versus atomic position.  Specifically, we experimentally show that the interference of near-degenerate cavity modes leads to both a strong and \textit{tunable-range} interaction between Bose-Einstein condensates (BECs) trapped within the cavity.  We exploit the symmetry of a confocal cavity to measure the interaction between real BECs and their virtual images without unwanted contributions arising from the merger of real BECs.   Atom-atom coupling may be tuned from short range to long range.   This capability paves the way toward future explorations of exotic, strongly correlated systems such as quantum liquid crystals and driven-dissipative spin glasses.  
\end{abstract}

\maketitle

\begin{figure}[t!]
\includegraphics[scale=0.93]{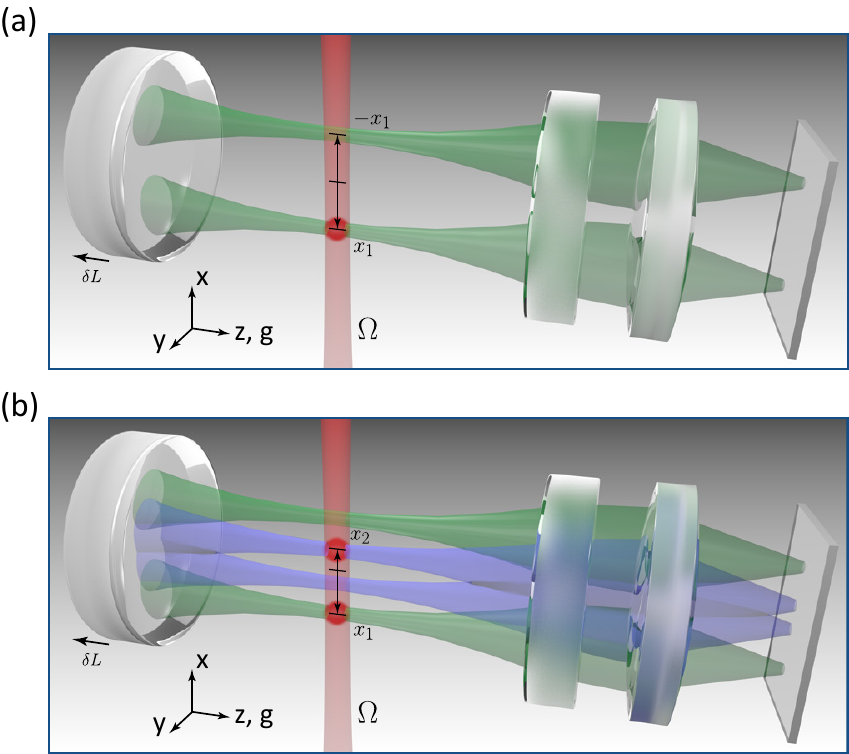}
\caption{\label{fig1} Sketches of experimental configurations employed. (a) A $\Rb{87}$ BEC (red circle) is trapped at the cavity waist $z=0$ at a location $x_1$ relative to the cavity center. The transverse pump beam (red beam) propagates along $\hat{x}$; the system undergoes a superradiant, self-organization phase transition above a critical field strength $\Omega_c$. Photons scattered off the BEC into the modes of the confocal cavity (green) create a virtual image (not shown) of the BEC at $-x_1$. The distance $\delta L$ indicates the tunable offset of the mirror from the confocal configuration. Emission of intracavity photons can either be sent to a single-photon counter, or be imaged onto an EMCCD camera to resolve the spatial structure of superradiant emission.  An absorption imaging laser for imaging BEC density travels along $\hat{y}$ (not shown). (b) Two $\Rb{87}$ BECs trapped at locations $x_1$ and $x_2$ on opposite sides of cavity center. Images of the two BECs are created at $-x_1$ and $-x_2$ (not shown).}
\end{figure}

\section{Introduction} \label{sec1}

Cavity QED provides strong light--matter coupling~\cite{Kimble1998}. For example, exotic nonlinear optical properties arise in cavity systems with atom-mediated photon-photon interactions~\cite{Peyronel2012}.   Realizations of topologically nontrivial states of interacting photons are within reach~\cite{Schine2016}.  Adiabatically eliminating the photonic field, rather than the atomic,  yields photon-mediated atom-atom interactions.  These interactions may be sufficiently strong to create novel quantum phases of matter~\cite{Ritsch2013}. Indeed, single and few-mode cavity QED in the optical domain have already provided demonstrations of supersolidity~\cite{Leonard2017,Leonard:2017wx} and exotic Mott physics~\cite{Landig2016,Klinder2015} in addition to   supermode-density-wave-polariton condensation~\cite{Kollar2017}.  Moreover, the driven-dissipative, open-quantum-system nature of cavity QED can change the character of quantum phase transitions, providing a new window into quantum nonequilibrium physics~\cite{Diehl2010,Sieberer2013}. 

An outstanding challenge has remained to create many-body cavity QED systems whose description requires physics beyond mean-field approximation.  Doing so enables, e.g., exploration of spin glass physics beyond the Sherrington-Kirkpatrick model where mean-field, replica-symmetry-breaking solutions may no longer hold~\cite{FisherHertz,Gopalakrishnan2011,Strack2011}, or quantum liquid crystals and intertwined orders such as those found in strongly correlated materials like high-T$_c$ superconductors~\cite{Fradkin2010,Fradkin2015,Gopalakrishnan2009,Gopalakrishnan2017}. More generally, strongly fluctuating, inhomogeneous (and frustrated) systems may organize in unexpected ways and the resulting surprises may lead to a deeper understanding of how quantum matter organizes.  A crucial limitation to exploring such physics using cavity QED stems from the fact that the single or few-mode cavities employed so far admit photon-mediated interactions that are all-to-all in coupling~\cite{Ritsch2013}.  The global (infinite-range) nature of these interactions necessarily implies that mean-field approximations are adequate to explain observed physics~\cite{altland2006condensed}.  However, it has been suggested that this challenge may be met either by employing networks of single-mode cavities~\cite{schmidt2013circuit,noh2016quantum}, using squeezed light to engineer interactions~\cite{zeytinoglu2017squeezed}, or by using a single multimode cavity~\cite{Gopalakrishnan2009,Gopalakrishnan2010,Gopalakrishnan2011}. 

This work presents a realization of a multimode cavity QED system and demonstrates that such a system does indeed provide strong, tunable, and local interactions among intracavity atoms. While no beyond-mean-field physics is yet explored, we show that the crucial ingredient of local interactions is present in the system, opening the road to future investigations where beyond-mean-field physics may be manifest.  

We measure the interaction range versus tunable parameters by manipulating the position of Bose-Einstein condensates (BECs) within the cavity. The symmetry of our confocal cavity is exploited to measure the interaction between real BECs and their virtual images without unwanted contributions arising from the merger of real BECs.  The experimental results are compared to theory, with good agreement. Furthermore, we show that the reduction in interaction range is accompanied by an increase in the effective atom-light coupling strength ($g_{\mathrm{eff}}$) and an emergence of a continuous translational symmetry in the plane transverse to the cavity axis.  

The paper is organized as follows. Section~\ref{interactionintro}  describes in general terms how tunable-range, photon-mediated interactions arise in a transversely pumped multimode cavity QED system undergoing a superradiant, self-organization transition.   Section~\ref{exp} then describes the cavity apparatus and BEC production and manipulation.  Section~\ref{results} presents the experimental results while Sec.~\ref{theorycomp} compares these to theory.  Section~\ref{sec:theory} discusses in greater detail the theoretical calculation of the photon-mediated atom-atom interaction.

\section{Photon-mediated interactions in a multimode cavity}~\label{interactionintro}
Atomic gases placed in transversely-pumped optical cavities have been shown to undergo a superradiant, self-organization transition arising from the competition between their free-particle dispersion and cavity-mediated interactions \cite{Domokos2002,Black2003,Baumann2010,Arnold2012,Kessler2014}.  Figure~\ref{fig1} shows examples of transversely pumped cavities.  For a pump laser red-detuned from the cavity resonance, atoms separated by a pump wavelength $\lambda$ along the cavity axis $\hat{z}$ constructively scatter pump photons into the cavity mode, leading to a buildup of intracavity light.  Conversely, scattering from atoms separated by $\lambda/2$ is suppressed. The resulting atomic light shift from the intracavity field creates an optical lattice potential that further localizes the atoms at integer-$\lambda$ separations.  The cavity may be interpreted to mediate a periodic, infinite-range interaction along $\hat{z}$ that lowers the energy of a $\lambda$-period atomic density wave.  Above a critical pump strength $\Omega = \Omega_c$, the cavity-mediated interaction energy of the density wave overcomes kinetic energy $2\mathcal{E}_\mrm{r}=h^2/m \lambda^2$ and the atoms self-organize into a $\lambda$-periodic pattern~\footnote{The  atomic recoil energy is $\mathcal{E}_\mrm{r}$.}.  In doing so, the atoms spontaneously choose to localize at either the even or odd antinodes of the standing wave. Interference between the pump and cavity beams means this even/odd choice is staggered along the pump direction and leads to a 2D checkerboard lattice in the $xz$-plane~\cite{Domokos2002,Asboth2005}. Concomitantly, the atoms superradiate into the cavity.  This second-order nonequilibrium phase transition is heralded both by a change in the atomic distribution~\cite{Baumann2010} and by an increase in cavity emission proportional to $N$~\cite{Black2003}.  The momentum distribution of the atoms may be detected in time-of-flight imaging, where Bragg peaks appear at wavevectors associated with the $\lambda$-period checkerboard lattice~\cite{Baumann2010,Kessler2014}. 

In conventional Fabry-Per\'{o}t cavities, i.e., those supporting a single TEM$_{0,0}$ mode near the pump frequency, the $(x,y)$ position dependence of the interaction energy between atoms follows the Gaussian profile $\Xi_{0,0}(x,y)$ of this mode.  The interaction energy vanishes at distances larger than the mode waist $w_0$. We now describe the explicit form of this interaction. Atoms at position $\mbf{x}$ coherently scatter pump photons into the cavity mode at a rate $\eta = g_0 \Omega\Xi_{0,0}(\mbf{x})/\Delta_a$, according to second-order perturbation theory, where $g_0$  is the single-atom atom-cavity coupling rate at $\mbf{x}=0$.  This expression is valid when the atomic excited state can be adiabatically eliminated from the dynamics for sufficiently large detuning $\Delta_a$ of the pump. A virtual photon may be exchanged between atoms within a time given by the inverse of the pump-cavity detuning $\Delta_c$. This virtual photon mediates an interaction given by (to second-order in perturbation theory)~\cite{Gopalakrishnan2009}
\begin{equation}
U(\mbf{x},\mbf{x}') = \frac{\eta(\mbf{x})\eta(\mbf{x}')}{\Delta_c}=\frac{g_0^2 \Omega^2\Xi_{0,0}(\mbf{x})\Xi_{0,0}(\mbf{x}')}{\Delta_a^2\Delta_c}.
\label{Ueff_single}
\end{equation}
As mentioned above, this interaction energy  smoothly vanishes versus distance for $\mbf{x}-\mbf{x}'$  larger than $w_0$. Atoms in gases much smaller than $w_0$ interact with a global, all-to-all coupling, up to the sinusoidal variation in $\hat{z}$ due to the standing-wave field modulation between the cavity mirrors.  

\begin{figure*}[!t]
\includegraphics[width=\textwidth]{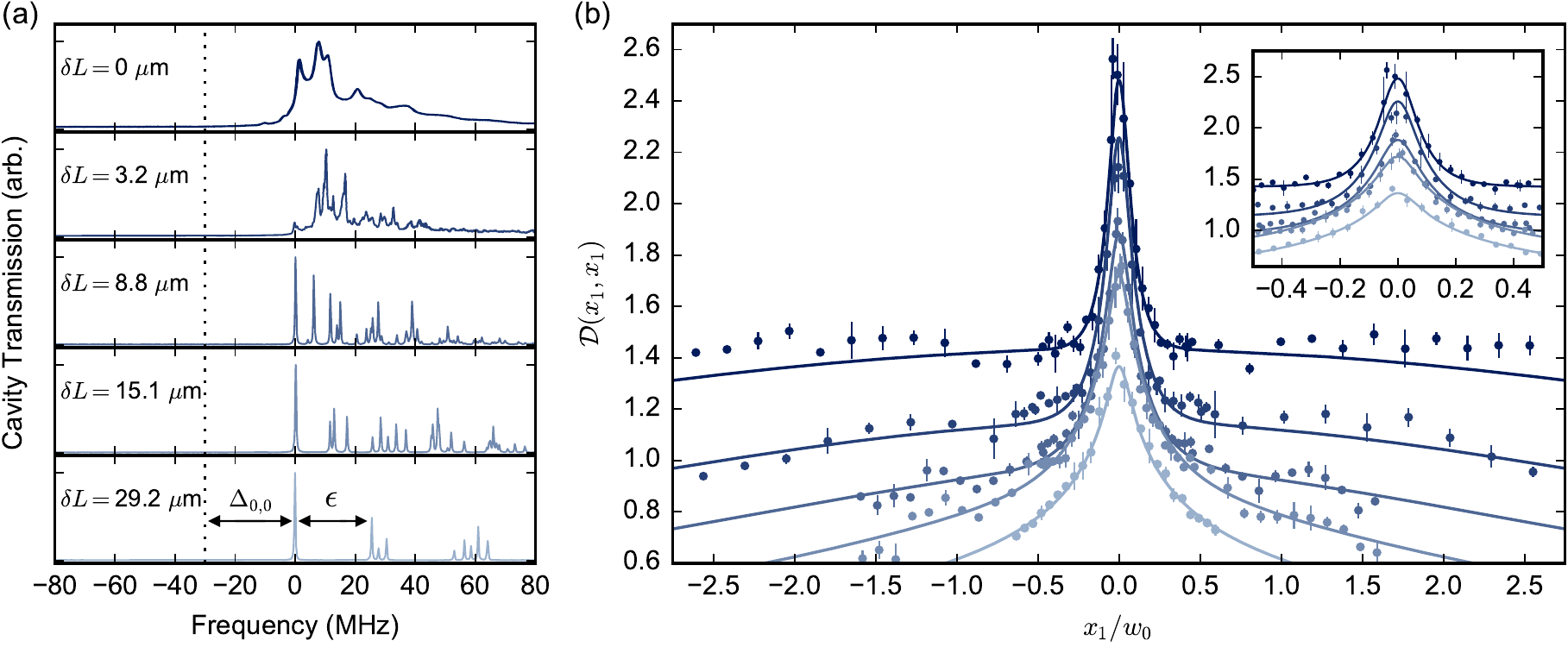}
\caption{\label{fig2} Interaction strengths $\mathcal{D}(x_1,x_1)$ versus position of a single BEC for various cavity lengths $L$. (a) Transmission spectra of the five cavities studied. The spacing between even-mode families is $\epsilon \approx 25$~MHz for $\delta L = 29.2$~$\mu$m.  Individual transverse modes are unresolvable at confocality ($\delta L = 0$~$\mu$m). The dotted vertical line indicates the frequency difference $\Delta_{0,0}=30$~MHz between the pump beam and the TEM$_{0,0}$ mode. (b) The spatial dependence of the interaction energy at $x_1$, also with $\Delta_{0,0} = 30$~MHz. The color of each trace corresponds to the cavity lengths presented in panel (a). The error bars represent one standard error in the mean over three runs. The solid lines are fits of Eq.~\ref{fit} to our data and neglect the effect of astigmatism and spherical aberrations on our cavity spectrum. A close-up of data near the cavity center is displayed in the inset.}
\end{figure*} 

This coupling need not be global in a multimode cavity, such as a confocal Fabry-Per\'{o}t resonator in which the cavity length $L$ equals the mirrors' radius of curvature $R$~\cite{siegman1986lasers}.   A multimode cavity can support several Hermite-Gaussian transverse modes at the same frequency, but with orthogonal mode functions $\Xi_{l,m}(\mbf{x})$.  $\Xi_{l,m}(\mbf{x})$ is the Hermite-Gauss function, describing the functional form of the TEM$_{l,m}$ mode at $(x,y)$ position $\mbf{x}$.   An atom scattering a pump photon into the cavity does so into a superposition of $\Xi_{l,m}$. The weights of the superposition are given by the mode strengths at the atomic position.  They are also given by any differences in detuning $\Delta_{l,m}$ between the $(l,m)$ modes and the pump due to residual differences $\epsilon$ in their mode frequencies.  The interaction energy then becomes
\begin{equation}
U(\mbf{x},\mbf{x}') = \frac{g_0^2 \Omega^2}{\Delta_a^2} \sum_{l,m} {\frac{\Xi_{l,m}(\mbf{x})\Xi_{l,m}(\mbf{x}')}{\Delta_{l,m}}}
    \mathcal{S}_{l,m}
\label{Ueff}
\end{equation}
When $\delta L = L- R$ is increased to move the system away from confocality, the Hermite-Gaussian modes of our near-confocal cavity exhibit a linear frequency dispersion with mode number:  $\Delta_{l,m} = \Delta_{0,0} + ( l + m ) \epsilon$. The factor $\mathcal{S}_{l,m}$, discussed in detail in Sec.~\ref{sec:theory}, accounts for the overlap between the atomic density wave and the photon mode  profiles along the cavity axis.  Due to the nature of confocal cavities, the sum over $(l,m)$ is restricted to $l+m$ being either odd or even~\cite{siegman1986lasers}.  Additional dispersion, present even at $\delta L = 0$, is due to mirror aberrations~\cite{Kollar2015}.  Ignoring aberrations for now,  we may rewrite the interaction in Eq.~\ref{Ueff} as
\begin{align}
U(\mbf{x},\mbf{x}') &= \frac{g_0^2 \Omega^2}{\Delta_a^2 \Delta_{0,0}} \mathcal{D}(\mbf{x},\mbf{x}') \label{Uint} \\
\mathcal{D}(\mbf{x},\mbf{x}') &= \sum_{l,m} {\frac{\Xi_{l,m}(\mbf{x})\Xi_{l,m}(\mbf{x}')}{1+(l+m)\epsilon/\Delta_{0,0}}} \mathcal{S}_{l,m},
\label{Ueff2}
\end{align}
where the spatial dependence of the interaction is encoded in the dimensionless function $\mathcal{D}(\mbf{x},\mbf{x}^\prime)$.  As we discuss in more detail below, the restriction to either odd or even modes means that this function can be thought of predominantly as containing two contributions, a direct interaction $\mathcal{D}_{\text{loc}}(\mbf{x},\mbf{x}^\prime)$ and its mirror image, 
$\mathcal{D}_{\text{loc}}(\mbf{x},-\mbf{x}^\prime)$.  As a result, we will see that the quantity appearing in Fig.~\ref{fig2}b, evaluated at equal positions $\mbf{x}=\mbf{x}^\prime$ contains two contributions: a broad background of self interaction providing a flat plateau from the direct term, and a sharp peak from the mirror term for small values of $x_1$.  The range of the cavity mediated interactions can be extracted from the width of this peak.

For an ideal cavity, supporting an infinite number of modes, there would be a delta-function interaction peak from $\mathcal{D}_{\text{loc}}(\mbf{x},-\mbf{x}^\prime)$ because the Hermite-Gaussian polynomials form a complete basis~\footnote{While this is true transverse to the cavity axis, it is mismatch in Gouy phase, not the destructive interference of Hermite-Gaussian polynomials, that would reduce the interaction range along the cavity axis.}, and the background contribution from $\mathcal{D}_{\text{loc}}(\mbf{x},\mbf{x}^\prime)$ would be entirely flat and nonzero. However, real cavities support only a finite number of modes, yielding a finite-range interaction: A photon is scattered into a wavepacket localized around the atom, and only atoms with overlapping polaritonic excitations---dressed atom-photon states---can interact.  

We note that there is a mirror image of the cavity mode focused through $-x_1$ and, for two atomic gases, through both $-x_1$ and $-x_2$.  These virtual images arise because confocal cavities only support modes of good parity at each degenerate resonance.  That is, the mode content alternates between all even or all odd modes every half free spectral range~\cite{siegman1986lasers}.  The fields at the real and virtual image locations are of the same (opposite) sign for cavities tuned to even (odd) modes resonances. We employ even mode configurations in this work.   The direct and mirror contributions can be seen in Fig.~\ref{fig1}a and b, which sketch the mode---supermode---that forms around each BEC for either one or two BECs in the cavity, respectively.   Each supermode is the mixture of bare cavity modes by the atomic dielectric~\cite{Wickenbrock2013,Kollar2017}.  The minimum waist of the supermode is as small as the width of the atomic gas if there are sufficiently many intracavity modes to create a compact superposition.

The form of $\mathcal{D}(\mbf{x},\mbf{x}')$ is set by the parameter $\epsilon/\Delta_{0,0}$, which may be experimentally controlled to tune the interaction range. The length scale of the range may be tuned between that of waist $w_0$ for a single-mode cavity to a small fraction of $w_0$ for a multimode cavity.  This is analogous to the \textit{phonon}-mediated interaction in ion traps, where large pump-detunings from resonances in the phonon spectrum generate shorter-ranged interactions~\cite{Porras2004,Kim2009,Britton2012}. 

To characterize the interaction profile $\mathcal{D}(\mbf{x},\mbf{x}')$, we use local measurements of the self-organization threshold for a small BEC. Using the expression for interactions in Eq.~\ref{Ueff2}, we may write a mean-field threshold condition for self-organization as
\begin{equation}
\frac{g_0^2 \Omega_c^2N^2}{\Delta_a^2 \Delta_{0,0}} \int \rho_{\mrm{TF}}(\mbf{x}) \mathcal{D}_c(\mbf{x},\mbf{x}')\rho_{\mrm{TF}}(\mbf{x}') \mrm{d}\mbf{x} \mrm{d}\mbf{x}' = 2N\mathcal{E}_\mrm{r},
\label{interactionvsKE}
\end{equation}
where $\rho_{\mrm{TF}}(\mbf{x})$ is the Thomas-Fermi density distribution of the BEC.  See Ref.~\cite{Gopalakrishnan2009,Gopalakrishnan2010} for the beyond-mean-field expression. For BEC radii much smaller than $w_0$, we may approximate the density by $\rho_{\mrm{TF}}(\mbf{x}) \approx \delta(\mbf{x}-x_1 \hat{\mbf{x}})$ to obtain an expression for $\mathcal{D}(x_1,x_1)$ at threshold,
\begin{equation}
\mathcal{D}_c(x_1,x_1) = \frac{2\mathcal{E}_\mrm{r}\Delta_a^2 \Delta_{0,0}}{N g_0^2 \Omega_c^2(x_1)} = \frac{\Omega_0^2}{\Omega_c^2(x_1)},
\label{singlecloud}
\end{equation}
where $\Omega_0$ is the threshold Rabi frequency for a delta-function-width gas localized at the center of a single-mode cavity. Measuring $\Omega_c(x_1,x_2)$ for a pair of atomic gases located at $x_1$ and $x_2$ allows one to determine $\mathcal{D}_c(x_1,x_2)$ via the relation
\begin{multline}
\frac{\Omega_0^2}{\Omega_c^2(x_1,x_2)}=
U_2(x_1,x_2)\\\equiv
\mathcal{D}(x_1,x_1) + \mathcal{D}(x_2,x_2) + 2\mathcal{D}(x_1,x_2),
\label{doublecloud}
\end{multline}
where we have dropped the subscript on the $\mathcal{D}$'s for convenience here and below.  The value of $\Omega_c(x_1,x_2)$ at which the superradiant, self-organization transition occurs allows us to measure the photon-mediated atom-atom interaction strength versus position through a closed form expression related to Eq.~\ref{Dloc} described in Sec.~\ref{results}.

When considering a pair of gases, we may exploit the mirror symmetry to cleanly measure the interactions between atoms in different gases without physically merging two real BECs.  That is, we use the fact that atoms in one gas can overlap and thus interact with the image of the other gas. By avoiding overlap of the real gases, we avoid unwanted systematics due to the change in atomic density and mean-field energy from collisions as the two traps merge. 
From the standpoint of photon-mediated atom-atom interactions, atoms at $x_i$ interact with their virtual images at $-x_i$ just like dipoles near a dielectric can be thought of as interacting with their mirror images in classical electrodynamics~\cite{jackson1975classical}.  

For $x_1 = x_2$, and away from $x_1=0$ where $x_1$ approaches $-x_1$, $\mathcal{D}(x_1,x_1)$ assumes a nearly flat distribution.  $\mathcal{D}(x_1,x_1)$ begins to decay as  a Lorentzian  at a  distance given by $w_0\sqrt{2M^*}$, where $(M^*)^2$ is a measure of the effective number of modes coupled to the atoms,  see Sec.~\ref{sec:theory}.  This provides a translationally invariant interaction energy over a large distance in the $xy$-plane.  For example, this distance is $\sim$200~$\mu$m on either side of our near-confocal cavity, far larger than typical BEC widths.  
We now describe the  characterization of the strength and range of  cavity-photon-mediated interactions for various pump and cavity configurations.  

\begin{figure}[!t]
\includegraphics[width = \columnwidth]{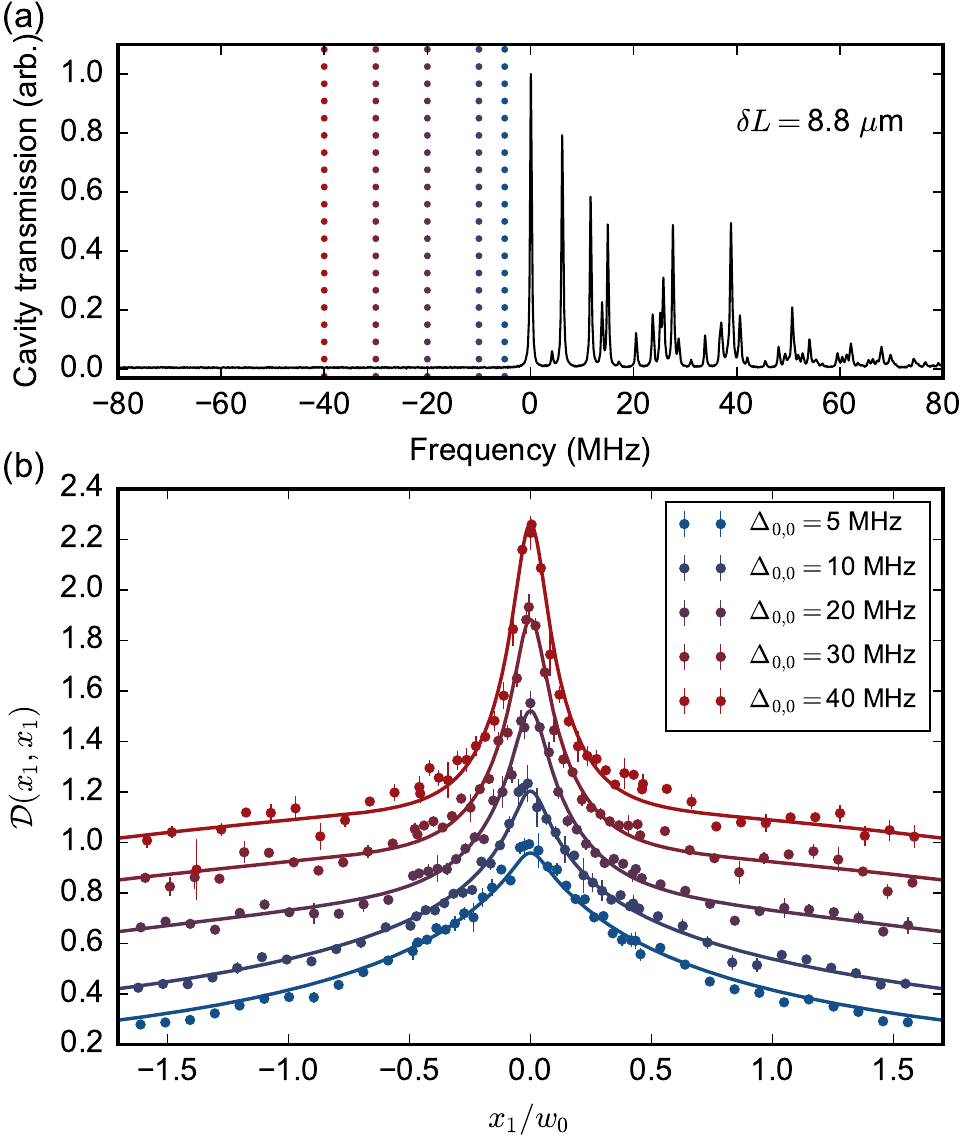}
\caption{\label{fig3} Interaction strength versus position of a single BEC for various pump-cavity detunings $\Delta_{0,0}$. (a) Transmission spectrum of the $\delta L = 8.8$~$\mu$m cavity presented in Fig.~\ref{fig2}a.  The dotted lines indicate the five  values of $\Delta_{0,0}$ at which the interaction energy of a single BEC was measured. The corresponding interaction energies are presented in (b). The solid lines are fits of Eq.~\ref{fit} to our data and neglect the effect of astigmatism and spherical aberrations in our cavity spectrum. The error bars represent one standard error in the mean over three runs.}
\end{figure}
 
 \begin{figure*}[!t]
\includegraphics[width=\textwidth]{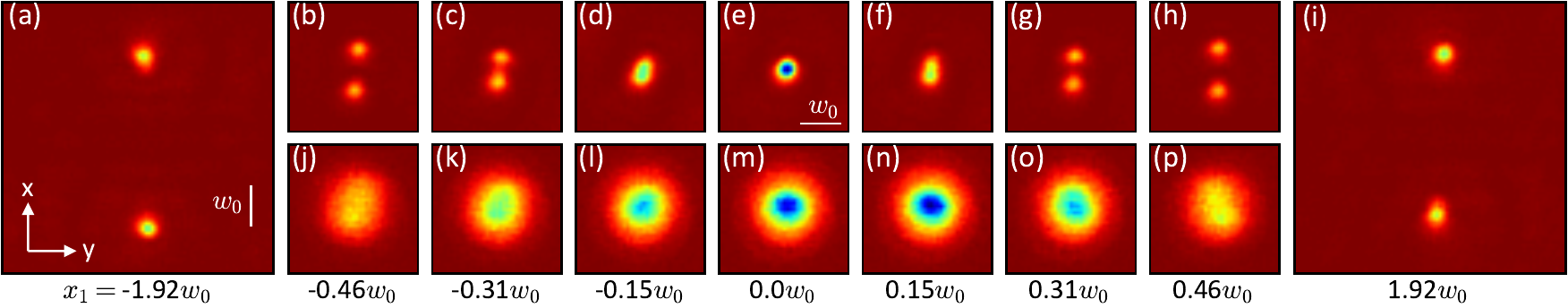}
\caption{\label{fig4} Superradiant emission into the cavity supermode above the self-organization threshold. (a)-(i) Spatial structure of superradiant emission into the modes of a near-confocal cavity ($\delta L = 0$~$\mu$m) as the BEC is translated from $x_1 = -1.92w_0$ to $1.92w_0$.  The two peaks merge at the center of the cavity, yielding a spot size smaller than the TEM$_{0,0}$ mode waist. Images (j)-(p) show superradiant emission for BECs in a single-mode cavity.  The BECs are in the same locations as in the above panels.  The cavity is set to $\delta L = 65.1$~$\mu$m, $\epsilon \approx 60$~MHz to achieve near-single-mode operation.  We see that the profiles are close to the width and shape of a TEM$_{0,0}$ mode. All images are plotted with identical length scales, including panels a and i, which have larger fields of view. Data are taken at a pump-cavity detuning of $\Delta_{0,0} = 20$~MHz. The white bars in (a) and (e) represent the length of the waist of the TEM$_{0,0}$ mode.}
\end{figure*}

\section{Experimental apparatus}\label{exp}
We investigate the behavior of photon-mediated interactions by trapping within an adjustable-length multimode optical cavity a BEC of $2.5(3) \times 10^5$ $\Rb{87}$ atoms in the $|F=1,m_F=-1 \rangle$ state.  See Ref.~\cite{Kollar2015} for BEC preparation procedure and Fig.~\ref{fig1} for illustration of experiment. The BEC is confined in a crossed optical dipole trap (ODT) formed by a pair of $1064$-nm laser beams propagating along $\hat{x}$ and $\hat{y}$ with waists of $40$~$\mu$m in the $xy$-plane and $80$~$\mu$m along  $\hat{z}$. The resulting trap frequencies of $(\omega_x,\omega_y,\omega_z) = 2 \pi \times [224(2),86(1),102(1)]$~Hz create a compact BEC with Thomas-Fermi radii $(R_x, R_y, R_z) = [4.0(1), 11.3(3), 8.3(1)]$~$\mu$m that are significantly smaller than the $w_0 = 35$~$\mu$m waist of the TEM$_{0,0}$ cavity mode. Acousto-optic deflectors (AODs) placed in the path of each ODT control the intensity and location of the ODTs, allowing us to translate the BEC to any point in the $xy$-plane with an uncertainty of $0.9~\mu$m. Some of the experiments discussed require two intracavity BECs that can be moved relative to one another.  We use dynamic trap shaping techniques~\cite{Bell2009} to split the BEC into two smaller BECs of $1.0(3) \times 10^5$ atoms each, with an imbalance uncertainty of $<$10\%.  These BECs may be separated in $\hat{x}$ by any relative distance using the AOD; see Fig.~\ref{fig1}(b).  Absorption imaging of the BECs is performed along $\hat{y}$ after a $15$-ms time-of-flight (TOF) to reveal the momentum distribution of either the single or double BECs.

The cavity is operated in a near-confocal regime in which the cavity length $L$ is set to be close to the mirrors' $R = 1$~cm radius of curvature. Due to astigmatism, there are two orthogonal radii of curvature that are slightly different. Because $L$ may only be set to match one radius at a time, the cavity is never perfectly confocal.  This contributes, along with spherical aberration, to the finite bandwidth (small spread) of modes seen in Fig.~\ref{fig2}(a) for $\delta L = 0$~$\mu$m~\cite{Kollar2015,Kollar2017}. The mode degeneracy is maximal when $L=R$, as shown in Fig.~\ref{fig2}(a)  for $\delta L = L-R=0$~$\mu$m. A slip-stick piezo attached to one of the mirrors allows us to change $L$ in situ~\cite{Kollar2015}. The frequency spacing $\epsilon$ between each family of transverse modes is controlled by $\delta L$, which provides tunability of mode density; see the transmission spectra in Fig.~\ref{fig2}(a). By family, we mean TEM$_{l,m}$ modes that satisfy $l + m = \text{const}$.  We have observed modes in cavity transmission with indices up to $l+m=300$.  This indicates that up to $\sim$$10^4$ modes are supported by the cavity near degeneracy.

The system with an atom at the field maximum of the TEM$_{0,0}$ mode has a single-atom cooperativity of $2.2(1)$, a vacuum Rabi splitting of $g_0 = 2\pi \times 1.47(3)$~MHz, and $\kappa = 2\pi \times 167(4)$~kHz~\cite{Kollar2015}. A laser propagating along $\hat{x}$ with Rabi frequency $\Omega$ pumps the BEC-cavity system near the even modes of the confocal cavity. The pump-cavity detuning $\Delta_{0,0}$ is defined as the difference in frequency between the pump and the TEM$_{0,0}$ mode. Where unclear, e.g., at small $\delta L$, the frequency of the TEM$_{0,0}$ mode is found by measuring the resonance frequency of a TEM$_{0,0}$ mode injected using a spatial light modulator~\cite{Papageorge2016}. The $\Delta_{0,0}$'s employed in this work are much larger than measured dispersive shifts at the atomic detuning of $\Delta_a = -102$~GHz. To achieve homogeneous pumping of the BEC and to minimize any perturbation to the BEC trap potentials, the transverse pump has a large waist ($1/e$ field radius) of $500$~$\mu$m. In contrast to the standing-wave pump configuration used in previous studies of cavity-induced self-organization~\cite{Baumann2010,Kessler2014}, we employ a running-wave pump~\cite{Arnold2012} in the data taken in Figs.~\ref{fig2},~\ref{fig3},~\ref{fig5}, and~\ref{fig7}.  This is done so as not to generate a lattice potential along $\hat{x}$ in the absence of intracavity light.  The absence of such a lattice leads to a simpler dependence of threshold pump power on the cavity mediated interaction:  For a standing wave, one must calculate the kinetic energy for atoms in the band-structure of the standing-wave lattice potential, and this means that pump power would appear on both sides of Eq.~(\ref{interactionvsKE}), making extraction of interaction strength less direct.  We do, however, use a standing-wave pump for the cavity output and atomic density images presented in Figs.~\ref{fig4},~\ref{fig5}, and~\ref{fig6} to avoid distortions due to atomic motion excited by running-wave pump.

\section{Measurements of cavity-induced interactions}\label{results}
We first measure the interaction energy of a single BEC as a function of its location in $\hat{x}$.  With the BEC trapped at a location $x_1$, the transverse pump power is linearly increased in time while the cavity emission is monitored on a single-photon counter.  A sharp increase in emission heralds the onset of the superradiant, self-organization transition and allows us to measure $\Omega_c(x_1)$, and consequently, using Eq.~(\ref{singlecloud}), extract the interaction strength $\mathcal{D}(x_1,x_1)$.

Figure~\ref{fig2}a shows the transmission spectra of the even modes in the five near-confocal cavities studied. Figure~\ref{fig2}b presents $\mathcal{D}(x_1,x_1)$ measured at a fixed pump detuning of $\Delta_{0,0} = 30$~MHz in the aforementioned cavities. For large values of $\delta L$, and consequently large $\epsilon$, the interaction strength would follow a single Gaussian decay as the BEC is moved further away from the center of the cavity.  This is because $\mathcal{D}(x_1,x_1)$ is following the mode profile of the TEM$_{0,0}$ mode in this near-single-mode cavity.  By contrast, as $\epsilon$ is reduced by shrinking $\delta L$, we observe a form with two components.  There is a flatter, more translationally-invariant background, falling off over a length scale $w$ coming from the self interaction of the gas. Furthermore, we observe the emergence of a prominent peak at the cavity center that decays over a much shorter range $\xi$ due to the interaction of the cloud with its mirror image. As $\epsilon$ becomes smaller, the scale $w$ grows and $\xi$ shrinks.  A similar behavior is observed for holding $\epsilon$ fixed but varying $\Delta_{0,0}$, as presented in Fig.~\ref{fig3}.

\begin{figure}[t!]
\includegraphics[width=\columnwidth]{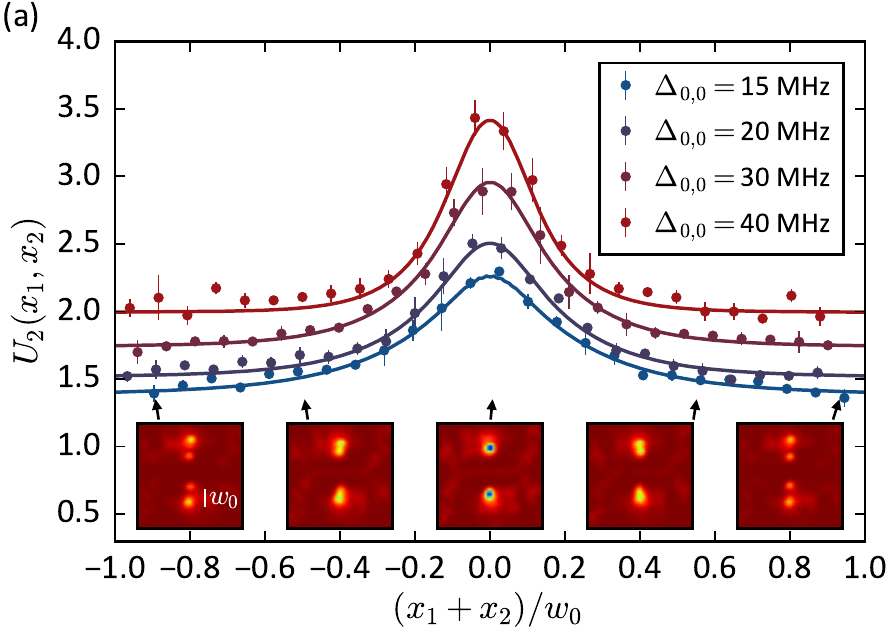}
\caption{\label{fig5} The local interactions versus position between a real  BEC  at $x_1$  and the virtual BEC at  $-x_2$ of a different real BEC at $x_2$.  The data were taken in the confocal configuration ($\delta L = 0$~$\mu$m) with the two BECs located on opposite sides of the cavity; see Fig.~\ref{fig1}b. Insets: The observed superradiant emission patterns for the data indicated. The white bar in panel b shows the length of the waist of the TEM$_{0,0}$ mode.  Error bars represent standard error.}
\end{figure}

The length scale $w$ of the background component reflects an overall envelope of the interactions, while the scale $\xi$ of the sharp peak reflects the interaction range.  Both the growth of $w$ and shrinking of $\xi$ can be understood as the result of superposing ever-larger high-order Hermite-Gaussian polynomials, allowing the interaction to both extend to larger distances and resolve finer features.
That we can measure the short-range interaction with only a single, compactly localized BEC is a consequence of the mirror symmetry inherent to confocal cavities. The interaction energy increases as the real atoms come near to their virtual images, even though there is only one real BEC present.  Viewed equivalently, as the two spots of the supermode begin to overlap,  the intracavity field magnitude increases, leading to a lower $\Omega_c$.

Images of the supermode can be directly observed in superradiant cavity emission patterns. Figures~\ref{fig4}a--i show examples in which the superradiant spots pass through each other. One cannot differentiate the spots from the picture alone, though from the orientation of the camera and apparatus, we know that the lower (upper) spot is the real image in panels a--d (f--i). The waists of the spots are smaller than that of a TEM$_{0,0}$ mode and their size at the object plane are similar to the BEC Thomas-Fermi radius, as expected. Their small size stands in stark contrast to the single-mode cavity's size shown in Fig.~\ref{fig4}(j)-(p):  The  superradiant emission pattern maintains its TEM$_{0,0}$ structure as the BEC is moved over the same distance in $\hat{x}$, only dimming as the gas nears the edge of the single Gaussian mode~\footnote{The slight elongation at $x_1 = \pm1.92 w_0$ may be due to the residual presence of higher modes in this near-single-mode cavity.}.

\begin{figure}[t!]
\includegraphics[width=0.75\columnwidth]{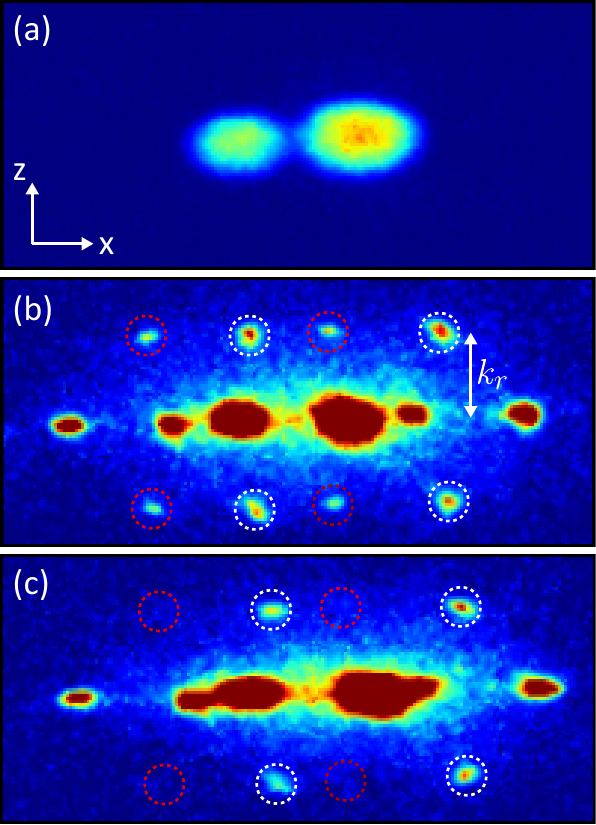}
\caption{\label{fig6} Absorption images in time-of-flight expansion of two intracavity BECs located on opposite sides of the cavity at $x_1$ and $x_2$.  The image is not purely of a momentum distribution due to the short time of flight.  The images show the contributions from each BEC along with the diffraction peaks from each gas.  (a) Time-of-flight expansion with no transverse pumping ($\Omega =0$). In this and the subsequent panels, the left BEC has 60\% fewer atoms than the one on the right.  (b) Time-of-flight expansion for a spacing of $x_1 = -x_2$; i.e., each real BEC spatially overlaps with the other BEC's virtual image.  The BECs self-organize at the same threshold pump Rabi frequency $\Omega = \Omega'_c$.  First-order Bragg peaks are visible for both the left BEC (red dashed circles) and right BEC (white dashed circles), heralding self-organization~\cite{Baumann2010}.  Additional diffraction peaks from the standing-wave pump beam are found to the left and right of each BEC.  (c) Separating the BECs from each other's virtual image by 32.4~$\mu$m, close to a cavity waist $w_0$, reduces the interaction energy, causing the small BEC to be unable to reach threshold at the same pump power as the larger BEC. ($\Omega = \Omega'_c$ is the same as in panel b.)  That is, the larger (smaller) BEC at right (left) exhibits (no) Bragg peaks, indicting (no) self-organization into a checkerboard lattice. The color scale has been increased in panels b and c with respect to panel a to make the Bragg peaks more visible. }
\end{figure}

We now present similar measurements of two identical intracavity BECs.  The BECs are located approximately $45$~$\mu$m from either side of the confocal cavity center, at $x_1$ and $x_2$ as illustrated in Fig.~\ref{fig1}(b). Each real BEC can then overlap and interact with the  nearby virtual image of the other BEC.  This is accomplished by moving $x_1$ and $x_2$ by the same amount in $\hat{x}$ while keeping $x_1 - x_2$ fixed.   Again, this allows us to probe the behavior of the photon-mediated interaction while avoiding any energy shifts due to density changes and $s$-wave collisions between the two BECs.
As shown in Figure~\ref{fig5}, we observe four distinct spots in the superradiant emission pattern---two at the BEC locations $(x_1,x_2)$ and two at the locations of their virtual images $(-x_1,-x_2)$. At $x_1+x_2 = 0$, the BECs merge with each other's virtual images, and we again observe a peak in $\mathcal{D}(x_1,x_2)$ arising from a  photon-mediated local interaction.  The sequence of BEC momenta observed by time-of-flight imaging in Fig.~\ref{fig6} further demonstrates how the  interaction energy of two nearby BECs can push a system above threshold:  A smaller BEC can undergo  self-organization at a lower threshold power when it is near a larger BEC than when it is far away.

\begin{figure}[t!]
\includegraphics[width=\columnwidth]{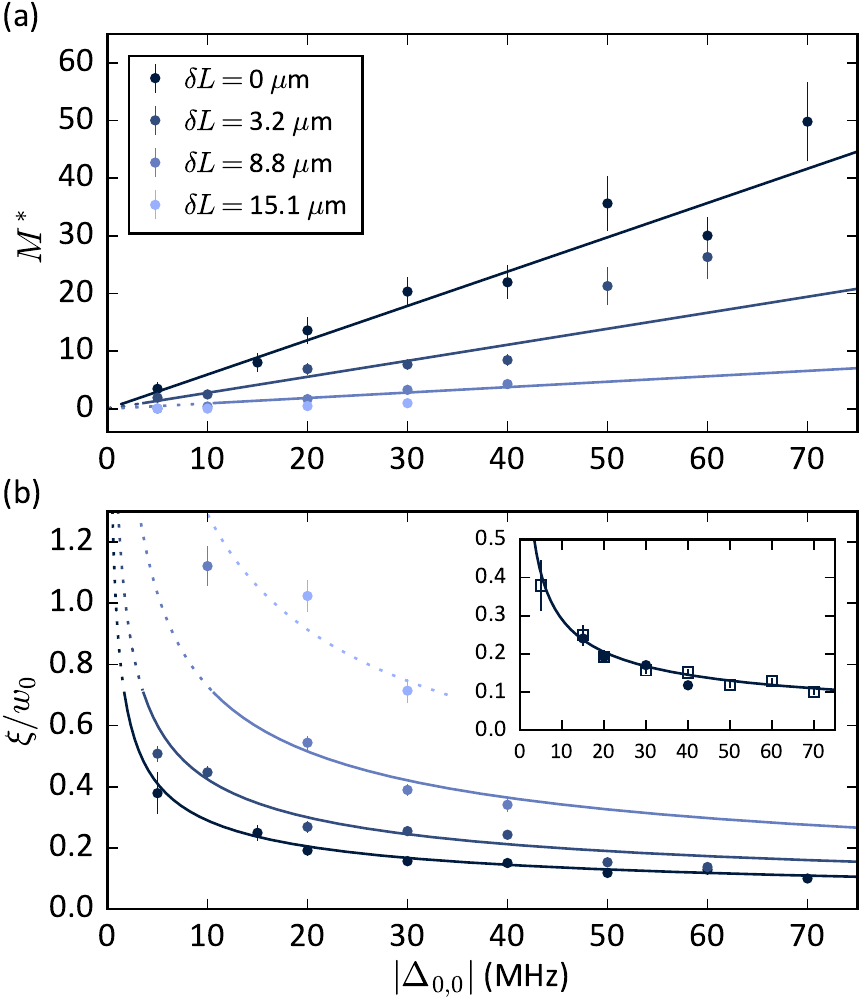}
\caption{\label{fig7} Tunability of the effective number of coupled modes, proportional to the square of  $M^*$, and interaction range $\xi$ versus  $\delta L$ and $\Delta_{0,0}$. (a) The dependence of $M^*$ on $\Delta_{0,0}$ for various cavity lengths $\delta L$. The solid lines are a fit to the theoretical expectation, $M^* \sim \Delta_{0,0}/\epsilon$. (b) The dependence of the interaction range $\xi/w_0 = 1/\sqrt{2M^*}$ on $\Delta_{0,0}$ inferred from the data in panel (a). We measure an interaction range of $\xi/w_0 = 0.09(1)$ at the largest value of $\Delta_{0,0}$ studied in the confocal configuration.  This is over an order-of-magnitude smaller than the TEM$_{0,0}$ waist $w_0$.  Solid lines are the same fits to the theoretical interaction profile as above.  The dashed lines are extensions of the fitted curve outside the regions of validity; i.e., where $M^*<1$.  Inset:  Agreement between interaction ranges extracted from the single-cloud (unfilled squares) and two-cloud (filled circles) datasets for $\delta L=0$~$\mu$m. The solid line is a fit to the single-cloud data.    All error bars represent standard error.}
\end{figure}

\begin{figure}[t!]
\includegraphics[width=\columnwidth]{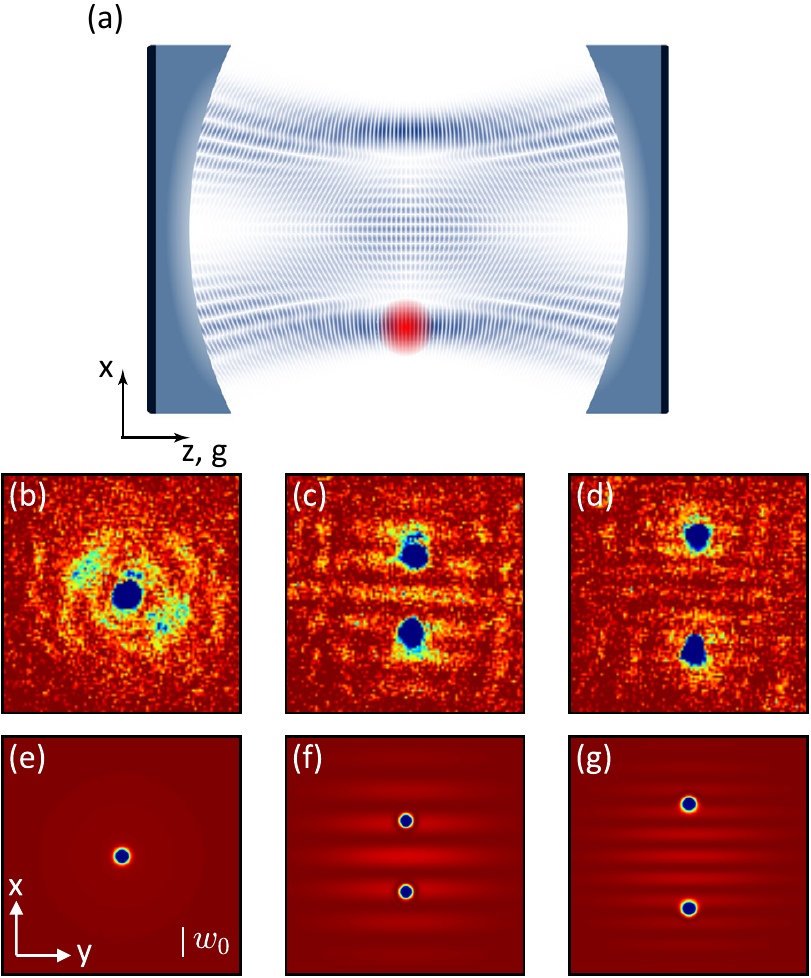}
\caption{\label{fig8} Manifestation of the non-local interaction $\mathcal{D}_{\mrm{non}}(\mbf{x},\mbf{x}')$. (a) Illustration of the hourglass structure in the supermode field (blue) when  populated by photons scattered into the confocal cavity from the BEC (red). The field displays a weak oscillatory behavior between the two spots at $x_1$ and $-x_1$. (b)--(d) Observed superradiant emission patterns for BECs placed at $x_1 = 0$~$\mu$m, $45.0$~$\mu$m, and $67.5$~$\mu$m, respectively. (e)--(g) Simulations of the intracavity field with the BEC at these locations.}
\end{figure}

\section{Measurement of interaction range}\label{theorycomp}

To extract the local interaction range from the decay of the peaks in interaction strength, we use a closed-form expression of $\mathcal{D}(\mbf{x},\mbf{x}')$---valid under the condition $\epsilon/\Delta_{0,0} \ll 1$---to fit the data in Figs.~\ref{fig2}(b),~\ref{fig3}(b), and~\ref{fig5}. See Sec.~\ref{sec:theory} for details.  This expression can be separated into three terms:
\begin{equation}
\mathcal{D}(\mbf{x},\mbf{x}') = \mathcal{D}_{\mrm{loc}}(\mbf{x},\mbf{x}') + \mathcal{D}_{\mrm{loc}}(\mbf{x},-\mbf{x}') + \mathcal{D}_{\mrm{non}}(\mbf{x},\mbf{x}')
\label{D},
\end{equation}
where $\mathcal{D}_{\mrm{loc}}(\mbf{x},\mbf{x}')$ is a local interaction between two atoms and $\mathcal{D}_{\mrm{loc}}(\mbf{x},-\mbf{x}')$ represents its corresponding atom-image interaction. The third term $\mathcal{D}_{\mrm{non}}(\mbf{x},\mbf{x}')$ is a weaker, non-local oscillatory interaction which will be discussed later.  

The local terms have  the form
\begin{equation}
    \label{Dloc}
    \mathcal{D}_{\text{loc}}(\vec{x},\vec{x}^\prime)
    =
    \frac{M^*}{4 \pi} 
    K_0\left( \frac{|\delta \mbf{x}| \sqrt{2M^*}}{w_0}
   \sqrt{1+\frac{\mbf{X}_{\mrm{cm}}^2 }{w_0^2 2M^* }}
    \right),
\end{equation}
where $K_0$ is the modified Bessel function of the second kind and falls off as an exponential for large $\delta \mbf{x} = \mbf{x}-\mbf{x}'$, the separation between the atoms.  The center-of-mass coordinate of the two atoms is $\mbf{X}_{\mrm{cm}} = (\mbf{x}+\mbf{x}')/2$. The strength and range of this interaction are controlled by the parameter $M^* = \Delta_{0,0}/\epsilon$.  The quantity $(M^*)^2$ may loosely be associated with the effective number of cavity modes that maximally couple in 2D to the BEC to form the supermode (see Sec.~\ref{defining_mstar}); $M^*$ is the number of modes that couple in 1D. We stress that the value of $M^*$  depends on the pump detuning and any aberration of the mirrors. Therefore one should not equate $(M^*)^2$ with the number of modes supported by the cavity near degeneracy, which is ${\sim}10^4$. Examining Eq.~\ref{Dloc}, we note that the interaction range $\xi = w_0/\sqrt{2M^*}$ decreases with increasing $M^*$.  A second length scale $w = \sqrt{2M^*}w_0$ controls the strength of this interaction as the pair of atoms is moved far from the cavity center.  Small $M^*$ dilutes the strength of this interaction versus distance from the cavity center:  $w$ may be interpreted as a measure of the degree to which the system is translationally symmetric.  $w$ diverges in an ideal confocal cavity as $\epsilon \rightarrow 0$, resulting in translationally-invariant interactions determined only by atomic separation $\delta \mbf{x}$, with no dependence on absolute position. 

We characterize the range of the local interactions in our cavity by fitting our data in Fig.~\ref{fig2} and \ref{fig3} to the theoretical model in Eq.~\ref{D}, while neglecting the weak non-local term $\mathcal{D}_{\mrm{non}}$. To account for the finite size of the BEC, $\mathcal{D}(x_1,x_1)$ is evaluated by numerically integrating over the BEC's Thomas-Fermi distribution $\rho_{\mrm{TF}}$ instead of a $\delta$-function:
\begin{equation} \label{fit}
\begin{split}
\mathcal{D}_{\text{eff}}(x_1,x_1;M^*) &= \iint \rho_{\mrm{TF}}(\mbf{x}-x_1)[ \mathcal{D}_{\mrm{loc}}(\mbf{x},\mbf{x}';M^*) + \\
& \mathcal{D}_{\mrm{loc}}(\mbf{x},-\mbf{x}';M^*)] \rho_{\mrm{TF}}(\mbf{x}'-x_1) \,d\mbf{x}\,d\mbf{x}'.
\end{split}
\end{equation}
We fit the above expression to our data using $M^*$ and an overall scale-factor as free fit parameters. Details of how this integral may be efficiently evaluated are given in Sec.~\ref{analytic_D}. The results of these  fits are shown as solid lines in Figs.~\ref{fig2}(b),~\ref{fig3}(b), and~\ref{fig5}. Extracted values of $M^*$ and the interaction range $\xi$ are presented in Fig.~\ref{fig7}(a) and (b) respectively, for several values of $\delta L$ and $\Delta_{0,0}$ using the single BEC configuration. Large values of $\Delta_{0,0}/\epsilon$ result in a more uniform coupling to transverse modes of the cavity and a shorter-ranged interaction.  With this control parameter, we can tune the interaction range to be as low as $\xi = 3.4(4)$~$\mu$m.  This is over an order-of-magnitude shorter than the range set by $w_0$ for a single-mode cavity.  Moreover, this close agreement between the data and fits for values of $M^*\agt 1$ highlights the validity of the theoretical model presented in Sec.~\ref{sec:theory}.   We note that we do not reliably infer $M^*$ for $M^* < 1$ because the closed-form expression in Eq.~\ref{Dloc} is only valid for $\epsilon \ll \Delta_{0,0}$.

We now turn our attention to the non-local interaction term in Eq.~\ref{D}, which displays oscillatory behavior of the form
\begin{equation}
\mathcal{D}_{\mrm{non}}(\mbf{x},\mbf{x}') \propto \cos{\left(\frac{\mbf{x}\cdot\mbf{x}'}{w_0^2}\right)}.
\end{equation}
As discussed below, the form of this term can be traced to the Gouy phase shifts of the bare-cavity modes~\cite{siegman1986lasers}.  While we cannot resolve the effects of this term in our interaction measurements, we do observe a weak signal in our images of superradiant cavity emission shown in Fig.~\ref{fig8}.  The cavity emission is recorded by imaging the plane containing the atoms onto the camera, so this emission records the light profile at the atom plane. For a single BEC at $\mbf{x_1}$, the image at position $\mbf{x}$ corresponds to $\mathcal{D}(\mbf{x},\mbf{x_1})$, and so the non-local term creates fringes in the cavity emission with a wavelength that becomes shorter as $x_1$ is increased, as shown in Fig.~\ref{fig8}b--d. The oscillatory behavior can most easily be understood by considering the ``hourglass" structure of confocal cavity modes.  While familiar ray-tracing representations of these modes depict the parallel and diagonal ``arms'' of the closed hourglass path~\cite{siegman1986lasers}, they do not account for interference of the paths.  A full calculation of the field of a confocal cavity supermode is shown in Fig.~\ref{fig8}a:  The parallel arms of the hourglass path create two spots at $x_1$ and $-x_1$, while the diagonal arms  interfere with each other to create fringes along $x$. A calculation of the superradiant intracavity field pattern shown in Fig.~\ref{fig8}e--g, using the theory presented in the next section, reveals a similar structure and is in qualitative agreement with our data.

\section{Theoretical model} \label{sec:theory}

\subsection{Hamiltonian and equations of motion}

To derive the atom-atom interaction, we start from a model of $N$ atoms in a condensate wavefunction $\Psi(\vec{r})$ interacting with cavity modes $\hat{a}_{\mu}$ by the Hamiltonian
\begin{multline}
\label{eq:Hamiltonian}
H= -\sum_\mu\Delta_\mu \hat{a}^\dagger_\mu \hat{a}_\mu  \\
+ N\int d^3\vec{r}
\Psi^\dagger(\vec{r})\left(-\frac{\nabla^2}{2m}+ V(\vec{r}) 
+U|\Psi(\vec{r})|^2\right)\Psi(\vec{r})\\
+\frac{N}{\Delta_a} \int d^3\vec{r}
\Psi^\dagger(\vec{r})|\hat{\phi}|^2
\Psi(\vec{r}),
\end{multline}
where for compactness we use $\mu = (l,m)$ to index the transverse modes of our cavity. The first term is the Hamiltonian of the cavity modes with detuning $\Delta_\mu=\Delta_{0,0}+(l+m)\epsilon$. The remainder is the standard Hamiltonian for a weakly interacting BEC with contact interactions of strength $U$ in an external trap $V(\vec{r})$ and with a Stark shift proportional to $\frac{1}{\Delta_a}$ due to the light in the cavity. This light field $\hat{\phi}$ consists of the running-wave pump and a sum over all cavity modes with their transverse and longitudinal spatial dependence
\begin{equation}
\hat{\phi}(\vec{r}) = \Omega e^{ikx} + g_0\sum_\mu \hat{a}_\mu
\Xi_\mu(\vec{r})\cos{[kz-\theta_\mu(z)]},
\label{totalLightField}
\end{equation}
where $\Omega$ is the pump Rabi frequency, $\Xi_\mu(\vec{r})$ is a Hermite-Gauss mode of the cavity and $\theta_\mu$ contains other contributions to the phase which vary slowly compared to $k z$. In particular, its dependence on $\mu=(l,m)$ is due to the Gouy phase term $(l+m) [\pi/4+\arctan(z/z_R)]$, where $z_R=L/2$ is the Rayleigh range, and this formula assumes $z$ is measured from the center of the cavity.  This term accounts for the fact that in order to have equal frequencies, a mode with higher order transverse structure must have a slower rate of change of longitudinal phase~\cite{siegman1986lasers}. The form of Eq.~(\ref{totalLightField}) results in a spatially varying single-photon Rabi frequency $g_0\Xi_\mu(\vec{r})/\Xi_{00}(0)$ for the mode $\mu$~\footnote{Note the sum is only over modes of the same parity; $l+m$ is even in this work.}. 

To study the location of threshold, we assume that most of the condensate is in the ground state, with a small fraction having a momentum kick from either scattering a photon from the pump into the cavity or vice-versa. Hence we write
\begin{multline}
\Psi(\vec{r})=Z(z-z_0)\left[\psi_0(\vec{r})+ \right. \\ \left.
\sqrt{2}\left(\psi_F(\vec{r})e^{ikx}+\psi_B^\ast(\vec{r})e^{-ikx}\right)\right],
\end{multline}
where $Z$ is an envelope function which describes the confinement of the gas in  $\hat{z}$, $\psi_0(\vec{r})$ is the wavefunction of the ground state of the gas in the transverse plane, and $\psi_{F(B)}$ is the wavefunction of the gas which has been scattered forward (backward) by scattering between the pump beam and the cavity modes.  Due to scattering into the cavity modes, these functions $\psi_{F,B}$ have a sinusoidal variation $kz$ along the cavity.  However, because of the different Gouy phase terms of different cavity modes, it is not a priori clear what phase the atomic density wave should take.  To allow the possibility of coupling to any cavity mode we further decompose the scattered atomic wavefunctions into two out-of-phase density waves
\begin{multline}
\psi_F(\vec{r})=\psi_{F1}(\vec{x})\cos{(kz-\theta_{0,0}(z_0))}\\+\psi_{F2}(\vec{x})\sin{(kz-\theta_{0,0}(z_0))},
\label{atomScatterTerm}
\end{multline}
and similarly for $\psi_B$. Here, $\vec{x}=(x,y)$ is the transverse coordinate vector, and $\psi_{F(1,2)}(\mbf{x})$ are now slowly varying envelope functions.  As we see below, different cavity modes couple preferentially to $\psi_{F1}$ or $\psi_{F2}$. For convenience, the phase offset $\theta_{0,0}(z_0)$ corresponding to the Gouy phase of the $(0,0)$ mode at the position of the atomic gas is introduced. We can now use Eq.~\ref{eq:Hamiltonian} to find the mean-field equations of motion for $\psi_{0,F,B}$ and $\alpha_\mu
\equiv\langle\hat{a}_\mu\rangle$. As the threshold is where the normal state $\alpha_\mu,\psi_{F,B}=0$ becomes unstable, we need only do this to leading order in these fields. 

To write equations in terms of only the transverse coordinates
$\mbf{x}$ we must perform the $z$  integral in Eq.~\ref{eq:Hamiltonian}.  This can be done straightforwardly in the limit where we assume $Z(z-z_0)$ has a width $\sigma_z$ and that $\lambda\ll\sigma_z\ll z_R$. The first inequality allows us to drop any terms oscillating at wavevector $k$; this imposes momentum conservation so that recoiling atoms pick up the difference of pump and cavity momenta.  The second condition means we can evaluate the slowly varying phase terms as being effectively constant over the width of the gas:  we can approximate $\theta_\mu(z) \simeq \theta_\mu(z_0) \equiv
\theta_{0,0}(z_0)+(l+m)\theta_0$ with $\theta_0 \equiv \pi/4 + \arctan(z/z_R)$.  In the linearized treatment, all relevant $z$ integrals involve the cross pump-cavity term causing scattering between at-rest atoms $\psi_0(\mbf{x})$ and the functions $\psi_{F,B}(\mbf{r})$.  We then find the $z$ integrals yield two possible values,
\begin{equation}
\mathcal{O}_\mu^i=\begin{cases} \cos{[(l+m)\theta_0]} & i=1 \\ \cos{[(l+m)\theta_0-\pi/2]} & i=2\end{cases} 
\end{equation}
For the equations of motion, we find
\begin{widetext}
\begin{align}
\label{eq:eom}
&i\partial_t\alpha_\mu =-(\Delta_\mu+i\kappa)\alpha_\mu -\frac{g_0^2 N}{2\Delta_a}
\int d\vec{x} |\psi_0(\vec{x})|^2\Xi_\mu(\vec{x})\Xi_\nu(\vec{x}) \alpha_\nu
-\frac{g_0 N\Omega}{\sqrt{2}\Delta_a}\int d\vec{x}\Xi_\mu(\vec{x})\psi_0(\vec{x})
\sum_{i=1,2}\left[\psi_{Fi}^\ast(\vec{x})+\psi_{Bi}(\vec{x})\right]\mathcal{O}_\mu^i \\
\label{eq:eom2}
&i\partial_t\psi_{Fi}(\vec{x}) =\left[-\frac{\nabla^2}{2m}+V(\vec{x})+2\omega_r+U|\psi_0(\vec{x})|^2\right]
\psi_{Fi}(\vec{x})+\frac{1}{2}U\psi_0^{\ast2}(\vec{x})\psi_{Bi}(\vec{x})-\frac{g_0\Omega}{\sqrt{2}\Delta_a}\sum_\mu\alpha_\mu^\ast\Xi_\mu(\vec{x})\psi_0(\vec{x}) \mathcal{O}_\mu^i,
\end{align}
\end{widetext}
where we have included photon loss $\kappa$ and $\omega_r$ is the recoil momentum $k^2/2m$. The ground state condensate has no linear perturbations, so at leading order we have:
\begin{equation}
i\partial_t\psi_0(\vec{x})=\left[-\frac{\nabla^2}{2m}+V(\vec{x})\right]\psi_0(\vec{x}),
\end{equation}
while $\psi_{Bi}(\vec{x})$ obeys an identical equation to Eq.~\ref{eq:eom2} with $F\leftrightarrow B$ and $\alpha_\mu\rightarrow\alpha_\mu^\ast$.  The ground state density profile is that of a Thomas-Fermi gas $\rho(\vec{x})=\rho_0[1-(x/x_0)^2-(y/y_0)^2]$, and so we have taken $\psi_0(\vec{x})$ to be real.

\subsection{Calculation of effective interaction $\mathcal{D}(\vec{x},\vec{x}^\prime)$}

We wish to study the effective photon-mediated atom-atom interaction. Since we expect the cavity field to reach a steady state on a timescale much faster than the atomic motion, we  adiabatically eliminate the photons by setting the time derivative in Eq.~\ref{eq:eom} to zero and solving for $\alpha_\mu$. We also neglect the corrections to the bare cavity modes caused by the ground state atomic gas; i.e., the term proportional to the integral of $|\psi_0(\vec{x})|^2$ in Eq.~\ref{eq:eom} is set to zero. Substituting this back into the equation of motion of the atomic condensate gives
\begin{multline}
\label{eq:atomeom}
i\partial_t\psi_{Fi}(\vec{x})=H_\mathrm{A}\psi_{Fi}(\vec{x}) 
+\frac{1}{2} U \psi_0(\vec{x})^2\psi_{Bi}(\vec{x}) \\
+\frac{g_0^2 \Omega^2 N}{2\Delta_a^2\Delta_{0,0}}\int d\vec{x}^\prime \sum_{j=1,2}
\mathcal{D}_{ij}(\vec{x},\vec{x}^\prime)
\psi_0(\vec{x})\psi_0(\vec{x}^\prime) \\ 
 \times\Bigl[\psi_{Fj}(\vec{x}^\prime)
+\psi_{Bj}(\vec{x}^\prime)\Bigr],
\end{multline}
where we defined an atomic Hamiltonian,
\begin{equation}
H_\mathrm{A}=-\frac{\nabla^2}{2m}+2 \mathcal{E}_\mrm{r} + V(\vec{x}) +U\lvert \psi_0(\vec{x})\rvert^2,
\end{equation}
and the cavity-mediated interaction takes the form:
\begin{equation}
\label{eq:kernel}
\mathcal{D}_{ij}(\vec{x},\vec{x}^\prime)=\Delta_{0,0}\sum_\mu\frac{\Xi_\mu(\vec{x})\Xi_\mu(\vec{x}^\prime)}{\Delta_\mu+i\kappa}\mathcal{O}_\mu^i\mathcal{O}_\mu^j.
\end{equation}
To simplify further, we assume that the atoms are close enough to the cavity center that $\theta(z_0)\approx\pi/4$.  In this case one may see that as long as $l+m$ is even, either $\mathcal{O}^1_\mu=0$ or $\mathcal{O}^2_\mu=0$, so the interaction becomes diagonal, $\mathcal{D}_{ij}(\vec{x},\vec{x}^\prime) = \delta_{ij} \mathcal{D}_i(\vec{x},\vec{x}^\prime)$ .  Furthermore, using standard trigonometric identities we can reduce the expression to:
\begin{align}
\label{UeffDefinition}
\mathcal{D}_{i}(\vec{x},\vec{x}^\prime)&=\Delta_{0,0}\sum_{l,m}\frac{\Xi_{l,m}(\vec{x})\Xi_{l,m}(\vec{x}^\prime)}{\Delta_{l,m}+i\kappa} \mathcal{S}^{i}_{l,m}
\\
\mathcal{S}^{i=1,2}_{l,m} &=
 \frac{1}{2}\left[
    1 \pm \cos{[(l+m)\pi/2]}
    \right]\left[1+(-1)^{l+m}\right]
    .
\end{align} 
In writing this, we introduced the factor $\left[1+(-1)^{l+m}\right]$ into $\mathcal{S}_{l,m}$ so that the  sum in Eq.~(\ref{UeffDefinition})  is now over all modes. This extra factor serves to cancel odd modes.  This rewriting will enable us below to make use of known expressions for sums of Gauss-Hermite functions multiplied by phase factors, $\exp[i \varphi(l+m)]$.
As a reminder, the detuning in the denominator takes the form $\Delta_{l,m} = \Delta_{0,0} + \epsilon(l+m)$

This term $\mathcal{D}_i(\vec{x},\vec{x}^\prime)$ is the expression given in Eq.~(\ref{Ueff2}), the interaction between atoms at different points $\vec{x}$ and $\vec{x}^\prime$ due to the cavity modes (except that in Eq.~(\ref{Ueff2}) we neglected cavity loss). Again, an identical equation to Eq.~\ref{eq:atomeom} holds for $\psi_B(\vec{x})$ with $F\leftrightarrow B$ and $\mathcal{D}(\vec{x},\vec{x}^\prime)$ replaced with its complex conjugate.

\subsection{Analytic forms of interaction near confocality}
\label{analytic_D}

In this section, we discuss those cases where it is possible to extract an analytic closed form for the interaction term,  Eq.~(\ref{UeffDefinition}). We are able to find expressions for both an ideally confocal system ($\epsilon = 0$) and a near-confocal cavity with $\epsilon \neq 0$.   Moreover, we show that restricting the number of  modes contributing to the interaction and including deviations from confocality affect the interaction profile similarly.  This connection allows us to identify, in Sec.~\ref{defining_mstar}, an effective number of modes $M^*$ that couple to the atoms.

If we first consider the ideal confocal case, $\epsilon=0$, the denominator in Eq.~(\ref{UeffDefinition}) becomes a constant, independent of $l,m$.  In this case,  we can make use of the harmonic oscillator Green's function,
\begin{multline}
G(\vec{x},\vec{x}^\prime,\alpha)=\sum_{l,m}\Xi_{l,m}(\vec{x})\Xi_{l,m}(\vec{x}^\prime) e^{-\alpha(l+m)} 
\\ = \frac{1}{\pi(1-e^{-2\alpha})}\exp{\left[
-\frac{\vec{x}^2+{\vec{x}^\prime}^2}{2 w_0^2\tanh(\alpha)}
+\frac{\vec{x}\cdot\vec{x}^\prime}{w_0^2\sinh(\alpha)}
\right]}.
\end{multline}
In terms of this, the interaction can be written as
\begin{multline}
\label{eq:greens}
\mathcal{D}_i(\vec{x},\vec{x}^\prime)=\frac{1}{4(1+i \tilde{\kappa})}
\lim_{\alpha \to 0}
\biggl[
G(\vec{x},\vec{x}^\prime,\alpha)+
G(\vec{x},-\vec{x}^\prime,\alpha)
\\ 
\pm\left(
G(\vec{x},\vec{x}^\prime,\alpha-i\frac{\pi}{2})+
G(\vec{x},-\vec{x}^\prime,\alpha-i\frac{\pi}{2})\right)\biggr],
\end{multline}
where $\tilde{\kappa}=\kappa/\Delta_{0,0}$ and we have made use of the relation
$G(\vec{x},\vec{x}^\prime,\alpha-i{\pi})=G(\vec{x},-\vec{x}^\prime,\alpha)$. 
If we then take the  limit of Eq.~\ref{eq:greens} for $\alpha\rightarrow 0$, we find the simple expression
\begin{multline}
\label{eq:confocal}
\mathcal{D}_i(\vec{x},\vec{x}^\prime)=
\frac{1}{4(1+i \tilde{\kappa})}
\biggl[
\delta\left(\frac{\vec{x}-\vec{x}^\prime}{w_0}\right)+
\delta\left(\frac{\vec{x}+\vec{x}^\prime}{w_0}\right)
\\ 
\pm\frac{1}{\pi}\cos\left(\frac{\vec{x}\cdot\vec{x}^\prime}{w_0^2}\right)\biggr]
\end{multline}
consisting of a local interaction between atoms, a local interaction between atoms and virtual atoms at their mirror image, and a non-translationally invariant oscillatory interaction.  

We can extend this result at confocality to find the interaction function in the limit of near confocality, where $\epsilon\ll\Delta_{0,0}$.  Defining $\tilde{\epsilon}=\epsilon/\Delta_{0,0}$ we rewrite the $l,m$ dependence of the denominator as an integral:  
\begin{multline}\label{eq:integral}
\mathcal{D}_i(\vec{x},\vec{x}^\prime) =
\sum_{l,m}
\frac{\Xi_{l,m}(\vec{x})\Xi_{l,m}(\vec{x}^\prime)}{1+(l+m) \tilde{\epsilon}+i\tilde{\kappa}} \mathcal{S}^i_{l,m}
\\
=\int_0^\infty d\tau e^{-\tau(1+i\tilde{\kappa})}
\sum_{l,m}
\Xi_{l,m}(\vec{x})\Xi_{l,m}(\vec{x}^\prime) \mathcal{S}^i_{l,m}
e^{-(l+m)(\tilde{\epsilon}\tau)} 
\\=\frac{1}{4} \int_0^\infty d\tau e^{-\tau(1+i\tilde{\kappa})}
\biggl[
G(\vec{x},\vec{x}^\prime,\epsilon\tau)+
G(\vec{x},-\vec{x}^\prime,\epsilon\tau)
\\ 
\pm\left(
G(\vec{x},\vec{x}^\prime,\epsilon\tau-i\frac{\pi}{2})+
G(\vec{x},-\vec{x}^\prime,\epsilon\tau-i\frac{\pi}{2})\right)\biggr].
\end{multline}
We may group the terms together as discussed in Eq.~(\ref{D}) to write
\begin{displaymath}
\mathcal{D}_i(\vec{x},\vec{x}^\prime) =
\mathcal{D}_{\text{loc}}(\vec{x},\vec{x}^\prime) +
\mathcal{D}_{\text{loc}}(\vec{x},-\vec{x}^\prime)
\pm\mathcal{D}_{\text{non}}(\vec{x},\vec{x}^\prime).
\end{displaymath}

The non-local contribution comes from the last two terms in Eq.~(\ref{eq:integral}).  By using the identities $\sinh(\theta - i \pi/2) = -i \cosh(\theta),\  \cosh(\theta-i\pi/2)=-i\sinh(\theta)$, we can write:
\begin{multline}
\mathcal{D}_{\text{non}}(\vec{x},\vec{x}^\prime)
=
\frac{1}{4} 
\int_0^\infty d\tau 
\frac{e^{-\tau(1+i\kaptil)}}{\pi(1+e^{-2\epstil \tau})}
\\\times
\exp\left[
-\frac{\vec{x}^2+{\vec{x}^\prime}^2}{2 w_0^2}\tanh(\epstil \tau)
\right]
2\cos\left[
\frac{\vec{x}\cdot\vec{x}^\prime}{w_0^2\cosh(\epstil \tau)}
\right].
\end{multline}
Because the first exponential suppresses contributions where $\tau \gg 1$, we may consider the small $\epstil$ behavior by making a small $\epstil \tau$ expansion, $\tanh(\epstil \tau) \simeq \epstil \tau$ and $\cosh(\epstil\tau)\simeq 1$ along with $1+e^{-2\epstil\tau} \simeq 2$.  The $\tau$ integral then becomes straightforward, yielding:
\begin{equation}
\label{eq:nonlocal_epsilon}
\mathcal{D}_{\text{non}}(\vec{x},\vec{x}^\prime)\simeq
\frac{\cos\left( \frac{\vec{x}\cdot\vec{x}^\prime}{w_0^2} \right)}{4 \pi \left[
    1+i \kaptil+
    \epstil\left(\frac{\vec{x}^2+{\vec{x}^\prime}^2}{2 w_0^2}\right) \right]}.
\end{equation}

For the local terms, a similar expansion for small $\epstil \tau$ is possible, however here we must note the prefactor involves $1-e^{-2 \epstil \tau} \simeq 2 \epstil \tau$.  We thus find:
\begin{multline}
\mathcal{D}_{\text{loc}}(\vec{x},\vec{x}^\prime)
=
\frac{1}{4}
\int_0^\infty d\tau 
\frac{e^{ - \tau(1+i \kaptil)}}{2 \pi \epstil \tau}
\\
\times\exp\left[
    - \frac{\epstil \tau}{2}
    \left(\frac{\vec{x}+{\vec{x}^\prime}}{2w_0}\right)^2
    - \frac{2}{\epstil \tau}
    \left(\frac{\vec{x}-{\vec{x}^\prime}}{2w_0}\right)^2
\right].
\end{multline}
The $\tau$ integral here can be shown to produce a modified Bessel function of the second kind, i.e.
\begin{multline}
    \mathcal{D}_{\text{loc}}(\vec{x},\vec{x}^\prime)
    =
    \frac{1}{4 \pi \epstil} \\ \times
    K_0\left( \sqrt{\frac{2}{\epstil}}
   \left|\frac{\mbf{x}-\mbf{x}^\prime}{w_0} \right|
   \sqrt{1+i \kaptil +\frac{\epstil}{2}
    \left(\frac{\mbf{x}+\mbf{x}^\prime}{2 w_0} \right)^2}
    \right).
\end{multline}

Because the Bessel function diverges at zero argument, it is crucial to consider the smoothed version of this function when comparing $\mathcal{D}_{\text{eff}}(x_1,x_1)$ in Eq.~(\ref{fit}) to experimental results.  In doing this, we may note that the two terms in the argument of the Bessel function have very different dependence on coordinates.  The first term depends strongly on the separation, with a characteristic length scale $w_0 \sqrt{\epstil/2}$, while the second term (inside the square root) has a much weaker dependence, with a characteristic length scale $w_0\sqrt{2/\epstil} \gg w_0$.  In the smoothed function $\mathcal{D}_{\text{eff}}(x_1,x_1)$,
we integrate over Thomas-Fermi distributions of the atom cloud. Assuming the cloud width is small compared to the length scale $w_0\sqrt{2/\epstil}$, we may neglect any difference between $\mbf{x},\mbf{x}^\prime$ and $\mbf{x}_1$ when evaluating the term in the square root.  This leads to the expression:
\begin{multline*}
\mathcal{D}_{\text{eff}}(x_1,x_1) = \frac{1}{4\pi\epstil}
\iint d\mbf{x}\,d\mbf{x}'
\rho_{\mrm{TF}}(\mbf{x}-\mbf{x}_1)
\rho_{\mrm{TF}}(\mbf{x^\prime}-\mbf{x}_1) \\
\Biggl[
   K_0\left( \sqrt{\frac{2}{\epstil}}
   \left|\frac{\mbf{x}-\mbf{x}^\prime}{w_0} \right|
   \sqrt{1+i \kaptil +\frac{\epstil}{2}
   \frac{x^2_1}{w^2_0} }
    \right)+\\
   K_0\left( \sqrt{\frac{2}{\epstil}}
   \left|\frac{\mbf{x}+\mbf{x}^\prime}{w_0} \right|
   \sqrt{1+i \kaptil} \right)
\Biggr].
\end{multline*}
Assuming symmetric clouds, this can further be simplified by suitable changes of variables to put it into the form of a convolution,
\begin{multline}
\label{Deffconvolution}
\mathcal{D}_{\text{eff}}(x_1,x_1) = \frac{1}{4\pi\epstil}
\int d\mbf{z} \rho_2(\mbf{z})\\
\Biggl[
   K_0\left( \sqrt{\frac{2}{\epstil}}
   \left|\frac{\mbf{z}}{w_0} \right|
   \sqrt{1+i \kaptil +\frac{\epstil}{2}
    \frac{x^2_1}{w^2_0}}
    \right)+\\
   K_0\left( \sqrt{\frac{2}{\epstil}}
   \left|\frac{2\mbf{x}_1-\mbf{z}}{w_0} \right|
   \sqrt{1+i \kaptil} \right)
\Biggr],
\end{multline}
where
\begin{math}
\rho_{2}(\mbf{z})
=\int d\mbf{y}
\rho_{\mrm{TF}}(\mbf{y})
\rho_{\mrm{TF}}(\mbf{z}-\mbf{y})
\end{math}.  This is the procedure used in fitting Figs.~\ref{fig2} and~\ref{fig3}. 

Note that the first Bessel function in Eq.~(\ref{Deffconvolution}) describes the ``self'' interaction of the cloud, and its only dependence on $x_1$ is via the square root in the Bessel function, which ultimately leads to a slow fall off with length scale $w\equiv w_0 \sqrt{2/\epstil} \gg w_0$.  The second term is the mirror interaction, and falls of exponentially with $x_1$ with a length scale $\xi\equiv w_0\sqrt{\epstil/2} \ll w_0$.
To see this behavior more clearly, we can consider the analytic expressions that result if we replace $\rho_{\mrm{TF}}(\mbf{x})$ by a Gaussian of width $\sigma$.
In this case $\rho_{2}(\mbf{z})$ is a Gaussian with width $\sqrt{2}\sigma$.
For the first term, which we denote $\mathcal{D}_{\text{eff,self}}(x_1)$,  we may use the result:
\begin{displaymath}
    \int d\mbf{z} \frac{e^{-{z^2}/{4 \sigma^2}}}{4\pi \sigma^2} 
    K_0\left( 2 A \left|\mbf{z}\right|\right)
    = \frac{1}{2} \int d\tau \frac{e^{-\tau}}{\tau + 4 \sigma^2 A^2},
\end{displaymath}
which comes from an integral representation of the Bessel function and defining
$A= (1/2\xi)\sqrt{1+i \kaptil + x_1^2/w^2}$.  
In this expression, the quantity $A \sigma  > (\sigma/2\xi) \gg 1$, and thus for the typical values of $\tau$ that dominate the integral, we have $4 A^2 \sigma^2 \gg \tau$. We thus find the first part of Eq.~(\ref{Deffconvolution}) has the form:
\begin{equation}
    \label{Deffselfgauss}
    \mathcal{D}_{\text{eff,self}}(x_1) = \frac{(w_0/\sigma)^2}{16\pi (1+i\kaptil + x_1^2/w^2)}
\end{equation}
There is no such simple closed form for the image term.  However, using the same approach as above we can write the expression in the form
\begin{displaymath}
    \mathcal{D}_{\text{eff,img}}(x_1) 
    =
    \frac{1}{8\pi\epstil}
    \int d\tau \frac{\exp\left(-\tau - \frac{4 x_1^2 A^2}{\tau + 4 \sigma^2 A^2}\right)}{\tau + 4 \sigma^2 A^2},
\end{displaymath}
where now $A= (1/2\xi)\sqrt{1+i \kaptil}$.  We still have that $A\sigma \gg 1$, however the extra terms in the exponent means it is no longer always true that the integral is dominated by terms for which $\tau \ll 1$.  At large $x_1$, the saddle point of the integral occurs when $\tau \simeq 2 A x_1$, and so for large enough $x_1$ we have that the dominant contribution comes from values for which $\tau \gg A^2 \sigma^2$.  The crossover occurs when $x_1 \simeq \sigma^2/\xi$.  We thus have two asymptotic limits:
\begin{equation}
    \label{Deffmirrorgauss}
    \mathcal{D}_{\text{eff,img}}(x_1) 
    =
    \begin{cases}\displaystyle
    \frac{(w_0/\sigma)^2 e^{-x_1^2/\sigma^2}}{16\pi\epstil(1+i\kaptil)}
    &
    x_1 \ll \sigma^2/\xi
    \\\displaystyle
    \frac{1}{4\pi\epstil}
    K_0\left( \frac{2 x_1}{\xi}
    \sqrt{1+i \kaptil}
    \right)
    &
    x_1 \gg \sigma^2/\xi
    \end{cases}.
\end{equation}

\subsection{Relating the ratio of mode dispersion to mode detuning $\tilde{\epsilon}$ to the effective number of coupled modes $(M^*)^2$}
\label{defining_mstar}

In order to make precise the sense in which we regard a non-zero $\epstil = \epsilon/\Delta_{0,0}$ as corresponding to a finite mode cutoff, we discuss here the results for such a cutoff.  For simplicity we consider a ``square'' cutoff, where we remove all modes 
$\Xi_{l,m}(x)$ with either $l,m>M$.  This means we may write expressions in terms of the 1D Green's functions.  Neglecting non-local terms we have
\begin{multline}
  \mathcal{D}_{M}(\mbf{x},\mbf{x}^\prime)
  =
  \frac{1}{4} \biggl[
    \mathcal{G}_M^{1D}(x,x^\prime) 
    \mathcal{G}_M^{1D}(y,y^\prime) 
    \\+
    \mathcal{G}_M^{1D}(x,-x^\prime) 
    \mathcal{G}_M^{1D}(y,-y^\prime) 
    +
    \ldots
  \biggr].
\end{multline}
The 1D Green's functions with finite cutoff can be written in terms of the Christoffel-Darboux identity to give:
\begin{align}
  \mathcal{G}_M^{1D}(x,x^\prime) 
  &=
  \sum_{n=0}^M \Xi_n(x) \Xi_n(x^\prime)
  \nonumber\\&=
  \frac{\Xi_{M+1}(x) \Xi_{M}(x^\prime) - \Xi_{M}(x) \Xi_{M+1}(x^\prime)}{x-x^\prime}.
  \label{eq:cdb}
\end{align}
Here and throughout this section we measure all lengths in units of the cavity beam waist, i.e., $w_0 \equiv 1$. Using the 1D Green's functions we want to evaluate:
\begin{align*}
  \label{eq:3}
  \mathcal{D}_{M,\text{self}}(\mbf{x}_1) &= 
  \frac{1}{4} 
    \mathcal{G}_M^{1D}(x_1,x_1) 
    \mathcal{G}_M^{1D}(y_1,y_1) 
    \\
    \mathcal{D}_{M,\text{mirror}} (\mbf{x}_1) &= 
    \frac{1}{4} 
    \mathcal{G}_M^{1D}(x_1,-x_1) 
    \mathcal{G}_M^{1D}(y_1,-y_1). 
\end{align*}
In the following, we will find approximate forms for these terms at large $M$.  For simplicity, we assume $M$ is even; similar results occur for odd $M$, but with various sign changes in intermediate formulae.   To consider the behavior at large $M$, we make use of the Wenzel-Kramers-Brillouin (WKB) approximation for a Gauss-Hermite function:
\begin{align*}
  \Xi_n(x) &\simeq
  E_n(x)
  \cos\left[ S_n(x) - n \frac{\pi}{2}\right], \\
  S_n(x) &= \int_0^x dz \sqrt{2n+1-z^2}, \\
  E_n(x) &= \frac{1}{\sqrt[4]{\pi(2n+1-x^2)}}.
\end{align*}

The mirror term has a simple form as we may write
\begin{displaymath}
  \mathcal{G}_M^{1D}(x,-x) =
  \frac{\Xi_{M+1}(x) \Xi_{M}(x)}{x}.
\end{displaymath}
The $1/x$ factor means we need only focus on behavior at small $x$.  This means we can approximate the phase function $S_M(x) \simeq  \sqrt{2M+1}x$ and the envelope function as $E_M(x) \simeq 1/\sqrt[4]{2M+1}$, and so we find
\begin{displaymath}
  \mathcal{G}_M^{1D}(x,-x) \simeq
  \frac{\sin( 2 \sqrt{2M+1}x)}{2 x \sqrt{\pi(2M+1)}}
  = \frac{\text{sinc}(2 \sqrt{2M+1}x)}{\sqrt{\pi}}.
\end{displaymath}
Thus, we find that the mirror term describes a sharp peak with a width that scales as $1/\sqrt{2M+1}$.  Comparing this to the results at non-zero $\epsilon$, we can identify an ``effective'' mode number, $M^* = \Delta_{0,0}/\epsilon$, as parameterizing this finite peak width.

The self interaction term is more subtle.  We can first rewrite the Green's function at $x^\prime \to x$ in terms of derivatives
\begin{displaymath}
  \mathcal{G}_M^{1D}(x,x+0) 
  =
  \Xi_{M}(x) \Xi^\prime_{M+1}(x) -
  \Xi_{M+1}(x) \Xi^\prime_M(x)
\end{displaymath}
and then use the recurrence relation on Gauss-Hermite functions:
\begin{math}
  \Xi^\prime_M(x)
  =
  \sqrt{2M} \Xi_{M-1}(x) - x \Xi_M(x)
\end{math}
to obtain:
\begin{multline*}
  \mathcal{G}_M^{1D}(x,x) =
  \sqrt{2(M+1)} \Xi_{M}(x)^2
  \\-
  \sqrt{2M} \Xi_{M+1}(x) \Xi_{M-1}(x).
\end{multline*}
One may now use that for large $M$, we can neglect differences between the envelope functions, $E_M(x) \simeq E_{M\pm 1}(x)$,  and approximate $\sqrt{2M} \simeq \sqrt{2(M+1)} \simeq \sqrt{2M +1}$ in the prefactors to write:
\begin{multline*}
  \mathcal{G}_M^{1D}(x,x) =
  \frac{ \sqrt{2M+1} E_M(x)^2}{2}
  \\
  \biggl[
    \cos(2 S_M(x))-
    \cos\left(S_{M+1}(x) + S_{M-1}(x) \right)
    \\+ 
    1
    +\cos\left(S_{M+1}(x) - S_{M-1}(x)\right)
  \biggr].
\end{multline*}
If we consider that:
\begin{multline*}
  S_{M\pm 1}(x) =
  \int_0^x dz \sqrt{2M+1-z^2} \times 
  \\\times\left[
    1 \pm \frac{1}{2M+1 - z^2}
    + \mathcal{O}\left( M^{-2} \right)
  \right],
\end{multline*}
one may readily see that $S_{M+1}(x) +S_{M-1}(x) = 2 S_M(x)+\mathcal{O}(M^{-2})$, and so to leading order in $1/M$ we have:
\begin{align*}
  \mathcal{G}_M^{1D}(x,x) &=
  \frac{ \sqrt{2M+1} E_M(x)^2}{2}
  \biggl[
    1
    +\cos\left(\delta S_M(x)\right)
  \biggr]
  \\
  \delta S_M(x) & \equiv 
  \int_0^x \!\!\! dz \frac{2}{\sqrt{2M+1-z^2}}
  = 
  2\arcsin\left( \frac{x}{\sqrt{2M+1}}\right)
  .
\end{align*}
Finally, using double angle formulae gives the result
\begin{equation}
  \mathcal{G}_M^{1D}(x,x) =\sqrt{\frac{2M+1-x^2}{\pi}}.
\end{equation}
This shows the self interaction term gives a broad semicircular function. Its algebraic form does not match the finite $\epstil$ result, but we can again identify the width of this function, $\sqrt{2M+1}$ with the width of the non-zero $\epsilon$ self interaction, to again give the identification $M^* = \Delta_{0,0}/\epsilon$.

\section{Concluding discussion}

The ability to engineer tunable-range interactions among intracavity atoms and, equivalently,  high-$M^*$ systems, opens several research directions.  We conclude with a discussion of three such directions, one involving exotic spatial organization of superfluid atoms and two involving spin organization.

The nature of the superradiant, self-organization phase transition can differ in a multimode cavity. In the single-mode cavity, the behavior is governed by mean-field theory, due to the all-to-all coupling.   In contrast, the multimode cavity allows transverse variations of phase across the cavity.  The atomic gases studied in this paper are (purposefully) too small to allow such variations, but by combining much larger intracavity BECs with the confocal cavity, transverse phase variation becomes possible.  This has a number of consequences.

An immediate consequence of transverse phase variation is the possibility of topological defects and phase textures. This is because atoms are no longer constrained to organize with respect to the profile of a single mode, but may fluctuate between the Hermite-Gaussian profiles of the multiple degenerate modes.  Similar to classical systems like diblock copolymers and fluids undergoing Rayleigh-B\'{e}nard convection, the organization should exhibit wandering stripe-like (smectic) patterns of atoms~\cite{Gopalakrishnan2009,Gopalakrishnan2010}. 
The interaction length scale $\xi$ controls the minimum size of a patch of stripes pointing in the same direction with the same spatial period, while the envelope of the interactions $w$ controls the maximum size of an atomic gas that can fully couple to the cavity.  As a result, the number of patches in the 2D transverse profile is ${\sim}(M^*)^2$.  The fact that we can engineer systems with $M^*\gg 1$ means that such complex, superfluid smectic states are within reach~\cite{Gopalakrishnan2009,Gopalakrishnan2010}.  This opens the door to exploring analogs of the quantum liquid crystals found in strongly correlated electronic materials, such as cuprate and iron-based high-T$_c$ superconductors~\cite{Fradkin:2012jm}.  Controllability of quenched disorder and dimensionality using external optical dipole trap beams and speckle would provide unique ways to investigate the intertwined nature of the order---crystalline, superfluid, and even magnetic (see below)---found in these systems~\cite{Fradkin:2015co}.

A second consequence of transverse degrees of freedom is their effect on the universality class of the phase transition.  For a single-mode cavity, the all-to-all coupling means the phase transition---analogous to the Hepp-Lieb-Dicke transition~\cite{Ritsch2013}---falls within the mean-field-Ising universality class.
This second-order mean-field phase transition is expected to become weakly first-order as the number of degenerate modes increases~\cite{Gopalakrishnan2009,Gopalakrishnan2010}.  This occurs in a scenario akin to that of a quantum version of the Brazovskii transition known from classical liquid crystal physics~\cite{Brazovskii:wr,Hohenberg:1995gm}.  As one approaches the critical pump strength for the second-order transition,  soft modes emerge corresponding to long-wavelength transverse fluctuations.  The additional, beyond-mean-field contribution to the effective action that arises from these fluctuations drives the transition first order.  

In addition to modifying the universality class of the phase transition, the presence of soft transverse modes can be seen in other ways. The dispersion relation of such modes could be measured through established methods for observing dynamical susceptibilities~\cite{Landig:2015et}.    
Just like  x-ray diffraction patterns of classical liquid crystals are arc-shaped~\cite{chaikin2000principles}, signatures of this quantum liquid crystalline state might appear as arc-like Bragg diffraction peaks in time-of-flight measurements. Because of the small size of the atomic gases, no such patterns are seen in Fig.~\ref{fig6}, but may become apparent by expanding the size of the intracavity BEC.  This may easily be accomplished by lowering the optical dipole trap frequencies. 

In the current configuration, the cavity mediates interactions between atomic density-wave excitations.  One can also consider cavity-mediated interactions between atomic spins.  These can be engineered if the transverse pump lasers drive a Raman transition between atomic Zeeman states representing a pseudospin-1/2 system~\cite{Dimer:2007da,Kastoryano:2011hr,Gopalakrishnan2011}.  If the atoms are trapped at random positions  inside the cavity to realize quenched disorder, then the multiple modes of the cavity can in principle mediate frustrated spinful interactions resulting in a spin glass-like state~\cite{Gopalakrishnan2011,Strack2011}.   However, there is a subtlety regarding the effects of summing many cavity modes: in some geometries, the sum over cavity modes may yield a short range interaction, in which case the degenerate limit produces a short-range spin model.  However,  as we have shown in this paper, the Gouy phase naturally present for a confocal cavity also induces  a long-range sign-changing interaction $\mathcal{D}_{non}(\vec{x},\vec{x}^\prime) \sim \cos(\vec{x}\cdot\vec{x}^\prime/w_0^2)$.  Such an RKKY-like sign-changing interaction is exactly the ingredient needed to enable glassy physics~\cite{FisherHertz}.  The ability to tune the relative strengths  between this long-range interaction and the short-range interaction $\mathcal{D}_{loc}(\vec{x},\vec{x}^\prime)$ provides a unique means (outside of numerical simulation) to experimentally compare the dynamics of infinite-range spin glasses to those with short-range interactions.  While the former has an order  known to be described by  mean-field replica-symmetry breaking, the latter's order defies explication despite many decades of investigation~\cite{stein2013spin}.  Direct spin-state detection combined with repeatable atomic disorder from shot-to-shot will allow us to create, observe, and compare system replicas.  This may provide sufficient experimental information to discriminate among various theories of short-range spin glass order. 

Spin glasses may serve as models for neural networks. Realizing spin glasses would provide the means to create a neural network comprised of atomic spins serving as neurons, cavity modes serving as  synapses, and  photons within the modes serving as action potentials~\cite{Gopalakrishnan2011,Gopalakrishnan:2012cf}.  Wiring the network to implement a particular graphical combinatorial optimization problem simply involves placing the atoms in specific locations within the cavity modes.  This may be possible with optical tweezer arrays~\cite{Barredo:2016ea,Endres:2016fk}. The combination of local and non-local  interactions demonstrated here will enable the construction of a wide variety of graphical combinatorial optimization problems, not just those of a complete graph.  In this way, Hopfield associative memories~\cite{Gopalakrishnan2011,Gopalakrishnan:2012cf,Torggler:2017hw,Rotondo:2017vg} and coherent Ising machines~\cite{McMahon:2016fy,Inagaki:2016eb} may be implemented in the presence of quantum effects like spin entanglement and quantum criticality, providing a new route to quantum neuromorphic computation.

\begin{acknowledgments}

We thank Sarang Gopalakrishnan for insightful discussions.  We are grateful for funding support from the Army Research Office. K.~E.~B. and J.~K. acknowledge support from EPSRC program TOPNES (EP/I031014/1). J.~K. acknowledges support from the Leverhulme Trust (IAF-2014-025).

\end{acknowledgments}


%

\end{document}